\documentclass{article}
\usepackage{amsmath,amsfonts,amscd,eucal,latexsym,amssymb}
\usepackage{epsfig}   
\usepackage{epsfig}

\newlength{\Taille}

\def\spa{\hskip -3pt} 
\newsymbol \rest 1316
\def\z{\zeta}
\def\mbH{\mathbb{H}}

\def\bM{\mathbb{M}}
\def\mch{\mathcal{H}}
\def\mcs{\mathcal{S}}
\def\mci{\mathcal{I}}
\def\mcg{\mathcal{G}}
\def\mcf{\mathcal{F}}
\def\mcd{\mathcal{D}}
\def\mcl{\mathcal{L}}
\def\mce{\mathcal{E}}
\def\mco{\mathcal{O}}
\def\bz{\overline{\zeta}}
\def\bZ{\mathbb{Z}}

\def\bms{G_{BMS}}
\def\gbms{G_{BMS}}
\def\tgbms{\widetilde{G_{BMS}}}

\newcommand{\flechebas}[1]{
  \settoheight{\unitlength}{\mbox{$#1$}}
  \settowidth{\Taille}{\mbox{~${\scriptstyle #1}$}}
  \addtolength{\unitlength}{4ex}
  \begin{picture}(0,1)
    \put(0,1){\vector(0,-1){1}}
    \put(0,0.5){\makebox(0,0){${\scriptstyle #1}$ \hspace{\the\Taille}}}
  \end{picture}}
\newcommand{\flechehaut}[1]{
  \settoheight{\unitlength}{\mbox{$#1$}}
  \settowidth{\Taille}{\mbox{~${\scriptstyle #1}$}}
  \addtolength{\unitlength}{4ex}
  \begin{picture}(0,1)
    \put(0,0){\vector(0,1){1}}
    \put(0,0.5){\makebox(0,0){\hspace{\the\Taille}${\scriptstyle #1}$ }}
  \end{picture}}
\newcommand{\flechedroite}[1]{
  \settowidth{\unitlength}{\mbox{$#1$}}
  \settoheight{\Taille}{\mbox{${\scriptstyle #1}$}}
  \addtolength{\Taille}{1ex}
  \addtolength{\unitlength}{4ex}
  \raisebox{0.5ex}{
  \begin{picture}(1,0)
    \put(0,0){\vector(1,0){1}}
    \put(0.5,0){\makebox(0,0){${\scriptstyle #1}$ \vspace{\the\Taille}}}
  \end{picture}}}
\newcommand{\flechegauche}[1]{
  \settowidth{\unitlength}{\mbox{$#1$}}
  \settoheight{\Taille}{\mbox{${\scriptstyle #1}$}}
  \addtolength{\Taille}{1ex}
  \addtolength{\unitlength}{4ex}
  \raisebox{0.5ex}{
  \begin{picture}(1,0)
    \put(1,0){\vector(-1,0){1}}
    \put(0.5,0){\makebox(0,0){${\scriptstyle #1}$ \vspace{\the\Taille}}}
  \end{picture}}}

\newcommand{\bC}{\mathbb{C}}

\def\bS{{\mathbb S}}
\def\bR{{\mathbb R}}
\def\bN{{\mathbb N}}
\def\bSf{{\mathbb S}^2}  
\def\scri{\Im^+}         
\def\scrip{\Im^-}         
\def\tg{\tilde{g}}

\def\beq{\begin{eqnarray}}
\def\eeq{\end{eqnarray}}

\newcounter{proposition}[section]
\newcounter{theorem}[section]
\newcounter{lemma}[section]
\newcounter{definition}[section]
\newcounter{remark}[section]

\def\theproposition{\thesection.\arabic{proposition}}
\def\thetheorem{\thesection.\arabic{theorem}}
\def\thelemma{\thesection.\arabic{lemma}}
\def\thedefinition{\thesection.\arabic{definition}}
\def\theremark{\thesection.\arabic{remark}}

\def\s #1 {\section{#1}}

\def\ssa #1 {\ifhmode{\par}\fi\refstepcounter{subsection}
  \noindent {\bf\thesubsection}. {\em #1}.\quad
  \addcontentsline{toc}{subsection}{\protect\numberline{\thesubsection} #1}%
  }

\def\ssb #1 {\ifhmode{\par}\fi\refstepcounter{subsection}
  \noindent {\bf\thesubsection.} {\em #1.}\quad
  \addcontentsline{toc}{subsection}{\protect\numberline{\thesubsection} #1}%
  }

\def\proposizione {\ifhmode{\par}\fi\refstepcounter{proposition}
  \noindent {\bf Proposition \theproposition}. \quad}
\def\teorema {\ifhmode{\par}\fi\refstepcounter{theorem}
  \noindent {\bf Theorem \thetheorem}. \quad}
\def\lemma {\ifhmode{\par}\fi\refstepcounter{lemma}
  \noindent {\bf Lemma \thelemma}. \quad}
\def\definizione {\ifhmode{\par}\fi\refstepcounter{definition}
  \noindent {\bf Definition \thedefinition}. \quad}
\def\remark {\ifhmode{\par}\fi\refstepcounter{remark}
  \noindent {\bf Remark \theremark}. \quad}

\begin{document}

\par

\noindent{\Large\bf Free field theory at null infinity and white noise calculus: a BMS invariant dynamical system} \\
\par

\noindent {\large \textsl{Claudio Dappiaggi}$\,^{a,}$\footnote{{ email:
claudio.dappiaggi@pv.infn.it}},} \\
\par
\small
\noindent $^a$ Dipartimento di Fisica Nucleare e Teorica, Universit\`a di Pavia, via A.Bassi 6 I-27100 Pavia, Italy.\smallskip
\normalsize

\par

\vspace{.5cm}

\noindent {\bf Abstract}. {In the context of asymptotically flat spacetimes we exploit techniques proper
either of white noise analysis either of dynamical systems in order to develop the Lagrangian 
and the Hamiltonian approach to a BMS invariant field theory at null infinity. }

\section{Introduction}

The quest to understand, to clarify and, to a certain extent, also to develop more in detail field
theory over curved backgrounds both at a classical and at a quantum level strongly relied
in the last decade on the holographic principle. 

Originally introduced by 't Hooft in \cite{'thooft} to study black hole backgrounds and the related
information paradox,  it still lacks nowadays of a mathematically rigorous and universally accepted 
definition in a general framework. 
Nonetheless the most promising though rather heuristic and, at the same time, demanding statement 
is the following: any field theory living on a $D$-dimensional manifold $M$
-  possibly including gravity - can be described by means of a suitable second field theory living on a 
codimension one submanifold of $M$.

From an abstract point of view such assertion is rather counterintuitive and at the same time revolutionary
since it states that the degrees of freedom encoding the information of a physical system 
evolving on a given background can be stored on a 
properly chosen lower dimensional region. On a practical ground, instead, even though one is inclined
to believe in such a conjecture,  it is straightforward to realize that the above formulation does not 
provide any concrete mean or hint on how to effectively implement the holographic paradigm. 

Amending such lack has been one of the main guideline in theoretical high-energy physics and in mathematical physics 
research for the past few years; at present a cornerstone is represented by the widely accepted 
realization of  holography for field theories living in asymptotically anti-de-Sitter spacetimes {\it i.e.} solutions 
to Einstein's field equation with a negative cosmological constant. 
Such a breakthrough (see \cite{Aharony} for an old but still complete review), 
originates from the so-called AdS/CFT correspondence; it hypothesize the existence of a 1:1
correspondence between a type IIB superstring theory living in the bulk of  AdS$_5\times S^5$ and a $SU(N)$
super Yang-Mills field theory living on its conformal boundary. Nonetheless, in order to catch some glimpses of its
complexity, one should notice that, at a mathematical level, a rigorous and complete proof of the Maldacena conjecture 
is still only partially available thanks to an analysis of free field theory in asymptotically AdS spacetimes within the framework
of the algebraic formulation of quantum field theory \cite{Rehren,Duetsch}.

In this paper we will not directly address the above briefly discussed topos whereas we will 
focus on a related though different problem namely if it is also possible to realize the holographic 
principle for field theories living on backgrounds solutions to Einstein's field 
equations with vanishing cosmological constant. To be more precise, we will deal with those spacetimes which are 
asymptotically flat at future (or past) null infinity {\it i.e.}, in a sense better specified in the next section,
they admit a conformal completion; thus they can be endowed with a natural
notion of boundary representing, as in the AdS/CFT correspondence, the codimension one submanifold
where to encode the data from a bulk field theory. 

Since the prototype of such class of manifolds is Minkowski spacetime,  from a physical perspective, there is
a strong interest in such a line of research since one could hope to concretely exploit holography to enhance our
comprehension of quantum field theory over curved background and to
eventually shed some light on some unsolved puzzles of quantum field theory over a flat background.  
From a mere mathematical perspective we will instead show how the development of a classical field theory at null infinity, as 
an holographic image of a suitable bulk counterpart, shows a deep-rooted and a priori unpredictable connection with 
recently developed techniques of functional analysis such as white noise calculus.  
Often binded to play a somehow ancillary role in classical or quantum field theory (see \cite{Albeverio} for a 
recent application or a \cite{Kuo} and references therein for a more detailed analysis), this last mentioned framework
represents the natural and the main machinery underlying the concrete development of a mathematically rigorous
field theory at null infinity thus providing a further interesting motivation for this kind of research (see also \cite{Dappiaggi} for
a preliminary analysis leading in this direction).

Nonetheless ``holography in asymptotically flat spacetimes'', as a whole, is not in its childhood
since a series of different papers appeared in the last few years discussing the
problem from different points of view (see \cite{deboer} and the recent \cite{Mann} where 
the concept of holography is geometrically
intertwined with spatial infinity instead of null infinity). In particular this paper can be placed
along the lines of \cite{Arcioni,Arcioni2} where it has been first explored and stated that a 
concrete realization of 't Hooft proposal not only should be developed at null infinity but it also 
must necessarily deal with a throughout analysis of the Bondi-Metzner-Sachs 
(BMS) group. As we will carefully discuss in the next sections, this is an infinite dimensional preferred subgroup of the 
diffeomorphism group of future (or past) null infinity which can be endowed with the structure of a semidirect product
between the proper orthocronous component of the Lorentz group and the set of
smooth functions over the 2-sphere seen as an abelian group under addition. 

To summarize, the rationale we advocate is the following: it possible to holographically encode the data of a field
theory living on an asymptotically flat spacetime in a BMS invariant free field theory at null infinity.  At present,
within this respect, several progresses have been achieved; as a starting point  we have
exploited group theoretical techniques in \cite{Arcioni} to classify and to construct, 
by means of Mackey theory of induction, the irreducible (and unitary) representations 
of the BMS group and the related induced or canonical wave functions. From a physical perspective,
these maps - discussed in section 3 - do represent the set of all possible dynamically allowed configurations
of the free fields of the theory and, together with their covariant counterpart, they 
also allow to fully characterize the dynamically allowed configurations for a BMS invariant field theory at null
infinity (see \cite{Dappiaggi,DMP}).  Furthermore these results have been recast in terms of an holographic
correspondence in \cite{DMP} where, exploiting the algebraic formulation of quantum field theory and the
related operator algebra techniques, several ``holographic theorems'' have been proved.  Between them, one
of the key achievement consisted in showing the existence of a 1:1 correspondence between 
a massless scalar field conformally coupled to gravity in any asymptotically flat, globally hyperbolic spacetime 
and a BMS invariant induced wave function intrinsically defined at null infinity\footnote{A further interesting result, 
though not directly connected with the aims of this paper, consists
on the identification  in \cite{DMP} of a preferred algebraic BMS invariant vacuum state at null infinity which 
can be suitably pulled-back in the bulk coinciding, in a flat background, with the usual Minkowski vacuum state.
Furthermore it  also been recently proved the uniqueness of such a state 
(see \cite{Moretti} for the demonstration and for a discussion of the main properties.)}

Nonetheless the whole approach advocated in \cite{DMP} suffers of two
main drawbacks; the first, which will not be discussed in this paper,
concerns bulk massive fields: even in globally hyperbolic spacetimes, there is no known way 
to coherently project a solution of the Klein-Gordon equation with $m\neq 0$ to a smooth function on 
future (or past) null infinity. Consequently a ``geometrical machinery'' projecting such bulk data on the
boundary is far from being completely constructed. The second 
deficiency, already sketched in the previous papers, consisted in a complete lack
of interactions in the boundary theories and, in particular, (Yang-Mills) gauge theories were not at all taken into account.
Such a deficency can be traced back to the overall approach that the infinite dimensional nature of the BMS group forced us to
take {\it i.e.} Wigner programme. This analysis provides a construction of the relevant free fields and of their
equations of motion without deriving the latter from a variational principle. Thus, at present, there is no real notion whether
the BMS field theory admits a Lagrangian formulation and more importantly an Hamiltonian formulation over a suitably
chosen symplectic space. From a point of view of interactions and in particular of gauge interactions, which are our ultimate
goal in an holographic analysis, it is known that it possible to rigorously construct the coupling between a free field theory and
gauge fields by means of a symplectic deformation of the free Hamiltonian system 
(see \cite{Weinstein,Guillemin} and in particular \cite{Landsman}).
Wishing ultimately to follow a similar road in the framework of BMS field theory, our aim in this manuscript 
is to cover the first part of the above sketched programme, namely to identify if an Hamiltonian system 
can be associated with a BMS free field, leaving to a future paper the symplectic deformation leading to 
gauge coupling. On an operative ground we will focus our attention on the ``working example'' of the massive and 
massless real scalar field and we will follow a two-step road. In the first part we will study
in detail the concept of a BMS covariant wave function which, although it appeared in the previously cited papers, has 
always played a sort of puppet role thus never being really carefully studied. 
To this avail, as we will outline mainly in section 4, we will require specific functional techniques such as white noise noise 
analysis which, thus, will be also covered with some care for sake of completeness. As a second step,
in this paper, we shall consider the BMS counterpart of the real scalar field equations of motion 
and we will solve the related inverse Lagrangian problem eventually also 
proving that an Hamiltonian description can as well be coherently formulated. 

As a side remark we wish also to point out that our attempt to construct a dynamical system at null infinity has been 
preceded by a now more than twenty five years old attempt by Ashtekar and Streubel in \cite{Ashtekar}.
In this paper the authors introduced a suitable Fr\'echet space as the configuration space for a smooth scalar
field living at null infinity and they exploit a theorem by Chernoff and Marsden from \cite{Chernoff} in order to construct
a BMS invariant symplectic phase space and an associated Hamiltonian phase space. Nonetheless the whole 
approach does not deal with the intrinsic ``Wigner-like'' formulation of a BMS invariant free field theory on null infinity 
and we shall comment further on this topic in the conclusions 

\vskip .4cm

\noindent\emph{Outline of the paper:} The paper is divided, beside the introduction and the conclusions,
in four main sections. 

In the next one we will briefly sketch the geometrical concept underlying an asymptotically flat spacetime at
future (or past) null infinity and we will introduce the key group-theoretical notion of the BMS group. 
Besides the recasting of already known concepts, the main issue of the whole section will be the introduction of an 
alternative and novel demonstration of a result partially exploited in \cite{DMP,Mc1} {\it i.e.}
we will prove that the BMS group and, more importantly, its abelian ideal are nuclear Lie groups.

Section \ref{tre} will briefly rephrase within the framework of fiber bundles, 
both the irreducible representations for the BMS group and the
set of induced wave functions. In particular we will discuss more in detail the 
key ``working'' examples of the massless
and the massive scalar field and we will develop an alternative way to introduce 
the Casimir invariant which plays the role of the mass for a field at null infinity.

In section \ref{quattro}, the main one of the paper, we will introduce the so-called covariant fields 
and we will show why and how the infinite dimensional nature of the BMS group 
forces us to introduce and fully exploit the powerful techniques of white noise calculus. 
Within this framework we will develop the key functional space where BMS field theory is defined 
in particular discussing the ``BMS counterpart'' of the Schwartz space of rapidly decreasing test functions and of 
distributions for a  BMS invariant field theory. 
At the end of the section we will also provide a demonstration of the Wigner programme in this specific framework
and the equations of motion for the BMS fields will be introduced as suitable operators on the above mentioned functional 
space. 

Eventually in the fifth and last section we will start from this latter result and we will
solve the inverse Lagrangian problem.  Exploiting an old analysis due to Gotay and Nester
on presymplectic Lagrangian systems we will also prove that the Lagrangian itself is almost regular and thus it is still 
possible to construct an associated Hamiltonian system.

\section{Asymptotically flat spacetimes and the BMS group}\label{seconda}
Throughout this paper we will refer to a {\em spacetime} as a four-dimensional 
smooth (Hausdorff second countable)  manifold $M$ equipped
with a Lorentzian metric $g$ assumed to be everywhere smooth; finally $M$ is 
supposed to be time orientable and time oriented. 
A {\em vacuum spacetime} is a spacetime satisfying vacuum Einstein equations.
 
We adopt the notion of {\em asymptotically flat at future null infinity} vacuum spacetime
presented in \cite{Wald} i.e. a smooth spacetime $(M,g)$ is called 
{\em asymptotically flat vacuum spacetime at null infinity} if there is 
a second smooth spacetime  $(\tilde{M},\tilde{g})$ such that $M$ turns out to be 
an open
submanifold of $\tilde{M}$ with boundary $\Im \subset \tilde M$.  $\Im$ is an 
embedded  submanifold of $\tilde M$ satisfying $\Im \cap \tilde{J^-}(M) = \emptyset$.
$(\tilde M,\tilde g)$ is required to be strongly causal in a neighborhood of 
$\Im$ and 
it must hold $\tilde{g}\spa\rest_M= \Omega^2 \spa\rest_M g\spa\rest_M$ where 
$\Omega \in C^\infty(\tilde M)$
is strictly positive on $M$. On $\Im$ one must have $\Omega =0$ and $d\Omega \neq 0$. 
Moreover, defining $n^a := \tg^{ab} \partial_b \Omega$, 
there must be a smooth function, $\omega$, defined in $\tilde M$ with $\omega >0$ on $M\cup \Im$, such that 
$\tilde{\nabla}_a (\omega^4 n^a)=0$ on $\Im$ and the integral lines of $\omega^{-1} n$ are complete on $\Im$.
Finally the topology of each set $\Im^\pm$ must be that of $\bS^2\times \bR$. 
$\Im$ is called {\em future null infinity} of $M$.\\
It is possible to make stronger the definition of asymptotically flat spacetime by requiring  asymptotic flatness
at both null infinity -- including the {\em past} null infinity $\scrip$ defined analogously to 
$\Im$ -- and {\em spatial} infinity, given by a special point in $\tilde M$ indicated by $i^0$. The complete definition 
is due to Ashtekar (see Chapter 11 in \cite{Wald} for a general discussion). We stress that the results presented in this work do not
require such a stronger definition: for the spacetimes we consider existence of $\Im$ is fully enough.

Considering an asymptotically flat spacetime,  the metric structures of $\scri$ are affected by a {\em gauge freedom}
due the possibility of changing the metric $\tg$ in a neighborhood of $\scri$
with a factor $\omega$ smooth and strictly positive. 
It corresponds to the freedom involved in transformations $\Omega \to \omega \Omega$ in a neighborhood of $\scri$.
The topology of $\scri$ (which is that of $\bR \times \bS^2$) as well as the differentiable structure 
 are not affected 
by the gauge freedom. 
Let us stress some features of this extent.
Fixing $\Omega$, $\scri$  turns out to be
the union of  future-oriented  integral lines of the field 
$n^a :=\tilde{g}^{ab}\tilde{\nabla}_b\Omega$.
This property is, in fact, invariant under gauge transformation, but the field $n$
depends on the gauge. 
For a fixed asymptotically flat vacuum spacetime $(M,g)$,
the manifold $\scri$ together with its degenerate metric $\tilde{h}$ induced by $\tg$ and
 the field $n$ on $\scri$
form a triple which, under gauge transformations $\Omega \to \omega \Omega$, transforms as
\beq
\scri \to \scri \:,\:\:\:\:\: \tilde{h} \to \omega^2 \tilde{h} \:,\:\:\:\:\: n \to \omega^{-1} n \label{gauge}\:.
\eeq
If $C$ denotes the class containing all of the triples $(\scri,\tilde{h}, n)$  transforming as in (\ref{gauge})
for a fixed
asymptotically flat vacuum spacetime $(M,g)$,
there is no general physical principle which allows one to select a preferred element in $C$.
Conversely, $C$ is {\em universal} for all asymptotically flat vacuum spacetimes in the following sense. 
If $C_1$ and $C_2$ are the classes of 
triples associated respectively to $(M_1,g_1)$
and $(M_2,g_2)$ there is a diffeomorphism $\gamma: \scri_1 \to \scri_2$ such that for suitable $(\scri_1,\tilde{h}_1, n_1)\in C_1$
and $(\scri_2,\tilde{h}_2, n_2)\in C_2$, 
\beq
\gamma(\scri_1) = \scri_2 \:,\:\:\:\:\: \gamma^* \tilde{h}_1=\tilde{h}_2 \:,\:\:\:\:\:\gamma^* n_1=n_2 \nonumber\:.
\eeq
The proof of this statement  relies on the following nontrivial result 
\cite{Wald}. For whatever asymptotically flat 
vacuum spacetime $(M,g)$ (either $(M_1,g_1)$  and $(M_2,g_2)$ in particular) and whatever initial 
choice for $\Omega_0$,
varying the latter with a judicious choice of the gauge $\omega$, 
one can always fix $\Omega := \omega \Omega_0$ in order that the metric $\tg$ associated with $\Omega$ satisfies
\beq
\tg\spa\rest_{\scri}  = -2du \:d\Omega +  d\Sigma_{\bS^2}(x_1,x_2)\:. \label{met}
\eeq
This formula uses the fact that in a neighborhood of $\scri$, $(u, \Omega,
x_1,x_2)$ define a meaningful coordinate system.
$ d\Sigma_{\bS^2}(x_1,x_2)$ is the standard metric on a unit $2$-sphere
(referred to arbitrarily fixed coordinates $x_1,x_2$) 
and $u \in \bR$ is nothing but an affine parameter along
the {\em complete} null geodesics forming $\scri$ itself with $n= \partial/\partial u$. In these coordinates $\scri$ is just the set of the points with
$u \in \bR$, $(x_1,x_2) \in \bS^2$ and, no-matter the initial spacetime $(M,g)$ 
(either $(M_1,g_1)$  and $(M_2,g_2)$ in particular), one has finally the triple 
$(\scri,\tilde{h}_B, n_B) := (\bR\times \bS^2, d\Sigma_{\bSf}, \partial/\partial u)$. \\

\definizione{ The {\bf Bondi-Metzner-Sachs (BMS) group}, $G_{BMS}$ 
\cite{Ashtekar,Penrose,Penrose2,Geroch}, is  the group of diffeomorphisms of 
$\gamma : \scri \to \scri$ 
which preserves the  universal structure of $\scri$, i.e. $(\gamma(\scri),\gamma^*\tilde{h}, \gamma^*n)$
differs from $(\scri,\tilde{h},n)$ at most by a gauge transformation (\ref{gauge}). }\\

Since it is convenient to provide an explicit representation of $G_{BMS}$ we 
need a suitable coordinate frame on $\scri$.   
Having fixed the triple $(\scri,\tilde{h}_B,n_B)$ one is still free to select an arbitrary coordinate frame 
on the sphere and, using the parameter $u$ of integral curves of $n_B$ to complete the coordinate system, one is
free to fix the origin of $u$ depending on $\z,\bz$ generally.
Taking advantage of stereographic projection one may adopt complex coordinates $(\z,\bz)$ on the (Riemann) sphere,
$\z= e^{i\phi}\cot(\vartheta/2)$,  $\phi,\vartheta$ being usual spherical 
 coordinates. \\
{\em Coordinates $(u,\z,\bz)$ on $\scri$ define a  {\bf  Bondi frame} when $(\z,\bz)\in \bC\times \bC$ are 
complex stereographic coordinates on $\bS^2$,
 $u\in \bR$  (with the origin fixed arbitrarily) is the 
 parameter of the integral curves of $n$ and $(\scri,\tilde{h},n)= (\scri,\tilde{h}_B,n_B)$.}\\ 
In this frame  the set $G_{BMS}$  is nothing but $SO(3,1)^\uparrow\times C^\infty(\bS^2)$, and
 $(\Lambda, f) \in SO(3,1)\sp\uparrow \times C^\infty(\bS^2)$ acts on $\scri$ as
 \cite{DMP}
\begin{eqnarray}
u &\to & u':= K_\Lambda(\z,\bz)(u + f(\z,\bz))\:,\label{u}\\
\z &\to & \z' :=\Lambda\z:= \frac{a_\Lambda\z + b_\Lambda}{c_\Lambda\z +d_\Lambda}\:, \:\:\:\:\:\:
\bz \: \to \: \bz' :=\Lambda\bz := \frac{\overline{a_\Lambda}\bz + \overline{b_\Lambda}}{\overline{c_\Lambda}\bz +\overline{d_\Lambda}}\:.
\label{z}
\end{eqnarray}
 \begin{eqnarray}
 K_\Lambda(\z,\bz) :=  \frac{(1+\left|\z\right|^2)}{\left|(a_\Lambda\z + b_\Lambda)\right|^2 + 
 +\left|(c_\Lambda\z +d_\Lambda)\right|^2}
\label{K}\:\:\; \mbox{and}\:\:
 \left[
\begin{array}{cc}
  a_\Lambda & b_\Lambda\\
  c_\Lambda & d_\Lambda 
\end{array}
\right] = \Pi^{-1}(\Lambda)\:.
\end{eqnarray}
$\Pi$ is the well-known surjective covering homomorphism $SL(2,\bC) \to 
SO(3,1)\sp\uparrow$. 
Thus the matrix  of coefficients $a_\Lambda, b_\Lambda, c_\Lambda, d_\Lambda$
is an arbitrary element of $SL(2,\bC)$ determined by $\Lambda$ up to an overall 
sign. However $K_\Lambda$\footnote{We adopt the convention of \cite{DMP} for the
analytic expression of $K_\Lambda(\z,\bz)$ which is slightly different from that of 
\cite{Mc1}. All results from this last cited paper will be adapted accordingly.} and the 
right hand sides of (\ref{z}) are manifestly independent from any choice of such a sign. 
\subsection{Group theoretical data}
Starting from (\ref{z}) and (\ref{K}), 
{\em in a fixed Bondi frame}, $G_{BMS}$ can be viewed as a regular semidirect 
product between $SO(3,1)\sp\uparrow$, the proper orthocronous subgroup of the Lorentz
group and the Abelian additive group $C^\infty(\bS^2)$ i.e.
$$G_{BMS}=SO(3,1)\sp\uparrow\ltimes C^\infty(\bS^2).$$
In particular, if $\odot$ denotes the  product in $G_{BMS}$, $\circ$ the 
composition of functions, $\cdot$
the pointwise product of scalar functions
and $\Lambda$ acts on $(\z,\bz)$ as said in the right-hand sides of (\ref{z}):
\begin{eqnarray}
K_{\Lambda'}(\Lambda(\z,\bz)) K_\Lambda(\z,\bz) &=& K_{\Lambda' \Lambda}(\z,\bz) \label{KK}\:.\\
(\Lambda',f') \odot (\Lambda,f) &=& \left(\Lambda' \Lambda,\: f + (K_{\Lambda^{-1}} \circ \Lambda)\cdot (f'\circ \Lambda)  \right)\:.
\label{product}
\end{eqnarray}\\
In the forthcoming discussion concerning the construction of field theories on
$\Im^+$, the $\bms$ group is going to play a key role and,
thus, it is necessary to better understand and characterize its structure. To
this avail, the first step consists of a carefull analysis of the $\bms$
subgroups and, in particular, of
$C^\infty(\bS^2)$ whose elements, in the physical literature, are usually referred to 
as {\bf supertranslations}. As a subgroup it is straightforward to
realize that it is an infinite-dimensional abelian ideal of $\gbms$; thus 
$C^\infty(\bS^2)$ as well as the full $\gbms$ group are not ordinary Lie groups.
Nonetheless, in the class of infinite-dimensional groups, they lie in a rather 
privileged class, the nuclear groups first introduced by Gelfand and Vilenkin
\cite{Gelfand}:\\

\definizione{\label{nuclear}
A group $G$ is a nuclear Lie group if it exists a
neighborhood of the unit element in $G$ which is homeomorphic to a neighborhood
of a countably Hilbert\footnote{A topological vector space over $\bC$ endowed
with a family of inner product norms $\left\{\left|\cdot\right|_p,\;
p\in\bN,\; p\geq1\right\}$ is called a \emph{countably Hilbert space} if it is
complete with respect to the topology induced by the norms.} nuclear space.}\\

In order to recognize if the set of supertranslations satisfies in a suitable
sense the above
definition, we shall make use of a construction for nuclear spaces often used in 
white noise calculus  \cite{Kuo,Hida,Hida2}:\\
\proposizione{
Let $\mch$ be any real separable Hilbert space with norm $\left|\left|,\right|\right|$
and let $A$ be any self-adjoint densely defined operator on $\mch$ such that it
exists an orthonormal base $\left\{e_i\right\}$ ($i\in\bN$) of $\mch$
satisfying the conditions:
\begin{enumerate}
\item $Ae_i=\lambda_i e_i\quad\forall i\in\bN$,
\item $1<\lambda_1\leq...\leq\lambda_n\leq...$
\item $\exists\alpha\in\bR_+$ such that
$\sum\limits_{j=1}^\infty\lambda_j^{-\alpha}\leq\infty$.
\end{enumerate}
If we introduce for any natural number $p$ the subspace of $\mch$
\beq\label{seq}
\mathcal{E}_p=\left\{\psi\in\mch\;\;\left|\left|\psi\right|\right|_p=\left|\left|
A^p\psi\right|\right|<\infty\right\},
\eeq
we can close each $\mathcal{E}_p$ to an Hilbert space with respect to the norm
$\left|\left|,\right|\right|_p$ and we can introduce the projective limit space
$\mathcal{E}=\bigcap\limits_p \mathcal{E}_p$. Let us equip $\mathcal{E}$ with
the projective limit topology $\tau_p$ i.e. an open neighborhood of the origin
in $\mathcal{E}$ is given by the choice $\epsilon>0$, $n\in\bN$ and by the set
$U_{\epsilon,n}=\left\{\psi\in\mch,\left|\left|\psi\right|\right|_n<\epsilon\right\}$.
Then a sequence $\left\{\psi_m\right\}_{m\in\bN}$ is said to converge to
$\psi\in\mathcal{E}$ iff it converges to $\psi$ in every Hilbert space
$\mathcal{E}_p$. The pair $\left(\mathcal{E},\tau_p\right)$ is metrizable and
complete thus it is a Fr\'echet space; furthermore the inclusion map
$\mathcal{E}_{p+\frac{\alpha}{2}}\hookrightarrow\mathcal{E}_p$ is
Hilbert-Schmidt, {\it i.e.} $\mathcal{E}$ is also a nuclear space.}\label{prima}\\

\teorema{\label{main1}
$C^\infty(\bS^2)$\footnote{From now on, within this paper, $C^\infty(\bS^2)$ will actually
refer to a set of equivalence classes. A smooth functions $\alpha(\z,\bz)$ will stand for 
a representative of the equivalence class $[\alpha(\z,\bz)]$ where $\alpha(\z,\bz)
\sim\widetilde\alpha(\z,\bz)$ if they differ for a function of zero measure with respect to the canonical
 measure over $\bS^2$.} is an infinite dimensional nuclear Lie group.}\\

\emph{Proof.}
Let us consider $\mch=L^2(\bS^2)$ i.e. the space of square integrable functions over
the two sphere with respect to the canonical volume element on $\bS^2$ and let us take
the operator $A=L^2+kI$ on $\mch$ where $k$ is any but fixed real number greater than $1$
and where $L^2$ is the angular momentum operator. Let us remember that the sphere
$\bS^2$ can be identified with the coset group $\frac{SO(3)}{SO(2)}$. Since $L^2$ is the
second order Casimir operator of $SO(3)$ it coincides with the Laplace Beltrami
operator on $\bS^2$ up to a factor -2. Thus, with
respect to the canonical local chart $(\theta,\varphi)$ on $\bS^2$, 
$2L^2=-\left(\frac{\partial^2}{\partial\theta^2}+\sin^2\theta\frac{\partial^2}{\partial
\varphi^2}\right)$. If we
choose the basis of spherical harmonics $\left\{Y_{lm}(\theta,\varphi), l\geq 0,
m=-l,...,l\right\}$, it is immediate to recognize either that $A$ is self-adjoint and
densely defined over $\mch$ either that 
$$AY_{lm}(\theta,\varphi)=\lambda_{lm}Y_{lm}(\theta,\varphi)=[l(l+1)+k]Y_{lm}
(\theta,\varphi)$$
Furthermore $1<\lambda_{00}<\lambda_{1-1}\leq\lambda_{10}<...$ and it holds that
$$\sum\limits_{l=0}^\infty\sum\limits_{m=-l}^l\lambda^{-\alpha}_{lm}=\sum\limits_{l=0}^\infty
(2l+1)[l(l+1)+k]^{-\alpha}<k^{-\alpha}+4\sum\limits_{l=1}^\infty
l^{-2\alpha+1},$$
which implies that the first sum is certainly convergent for
$\alpha>\frac{1}{2}$.

Thus the hypotheses of the previous proposition are satisfied and we may
construct for each $p\in\bN$ the space $\mathcal{E}_p$ as in
(\ref{seq}) with $A=L^2+kI$. Furthermore $\mathcal{E}_p\subset\mathcal{E}_q$ for
any $p>q\geq0$ and the inclusion map of $\mathcal{E}_{p+\frac{\alpha}{2}}$ in
$\mathcal{E}_{p}$ is given by the operator $A^{-\frac{\alpha}{2}}$ which
is an Hilbert-Schmidt operator with $\left|\left|
A^{-\frac{\alpha}{2}}\right|\right|^2_{HS}=\sum\limits_{l=0}^{\infty}(2l+1)
\left[l(l+1)+k\right]^{-\alpha}.$

Accordingly, given $\mathcal{E}=\bigcap\limits_p\mathcal{E}_p$ and $\tau_P$, the
induced limit topology defined in proposition \ref{prima}, the space 
$(\mathcal{E},\tau_p)$ is a nuclear (Fr\'echet) space. We now need to show that
$\mathcal{E}$ coincides with $C^\infty(\bS^2)$. Pick any smooth real function 
$\alpha(\z,\bz)$; by compactness of $\bS^2$, we can immediately conclude that
it lies also in $L^2(\bS^2)\equiv\mathcal{E}_0$. Furthermore since the operator
$A$ is, up to a factor $2$, minus the Laplacian operator plus a constant times the 
identity operator, it maps smooth functions in smooth functions.
It implies that $\alpha(\z,\bz)$ also lies in $\mathcal{E}_p$ for all $p$ and 
thus also in $\mathcal{E}$. Consequently $C^\infty(\bS^2)\subseteq\mathcal{E}$.

The converse is rather more difficult to prove. 
Referring to $\mathfrak{so}(3)$ as the Lie algebra of $SO(3)$, let us introduce
a set of generators $X_i\in\mathfrak{so}(3)$ with $i=1,...,3$; choosing a 
representation $\tilde T$ of $SO(3)$ in any but fixed Hilbert space 
$\mathcal{H}$, the corresponding Lie algebra representation $T$ is defined 
on the set of vectors $u\in\mch$ such that it exists the limit 
\begin{equation}\label{limit}
\lim\limits_{t\to 0}\frac{1}{t}\left[\tilde T\left(\exp(tX)\right)-I\right]u\doteq 
T(X)u,
\end{equation}
where $\exp(tX)$ is a one parameter subgroup of elements in $SO(3)$ obtained by
means of the exponential mapping. Furthermore, being $SO(3)$ compact, the
representation $\tilde T$ can be chosen unitary and consequently the operators 
$iT(X)$ are symmetric. 

Fix now $\mathcal{H}$ as $L^2\left(\frac{SO(3)}{SO(2)}\right)$ (where the internal product is defined with
respect to the unique $SO(3)$ invariant measure on $\frac{SO(3)}{SO(2)}$) and the 
unitary representation $T(g)$ as the right action i.e.
$$T(g)\psi(g^\prime)=\psi(g^\prime g).\quad\forall g\in
SO(3)\;\wedge\;\forall\psi(g)\in L^2\left(\frac{SO(3)}{SO(2)}\right)$$
Choose the generators $X_i\in\mathfrak{so}(3)$ in such a way that the Cartan 
metric tensor is diagonal; consequently $\Delta=\sum\limits_{i=1}^3 X^2_i$ is a
symmetric elliptic element of the enveloping algebra of $SO(3)$. Furthermore
$T(\Delta)=\sum\limits_{i=1}^3 T(X_i)^2$ is
essentially self adjoint on $L^2\left(\frac{SO(3)}{SO(2)}\right)$ since it coincides with
minus the angular momentum/Laplacian operator. Let us now introduce the Garding domain/subspace 
$D_G\subset\mathcal{H}=L^2(\bS^2)$ which is the linear subspace spanned by the
linear combinations of 
$$T(\varphi)\psi(g^\prime)=\int\limits_{SO(3)}d\mu^\prime(g)\varphi(g)T(g)\psi(g^\prime),$$
for all $\varphi(g)\in C^\infty\left(SO(3)\right)$ and for all 
$\psi(g^\prime)\in L^2\left(\frac{SO(3)}{SO(2)}\right)$. Here
$d\mu^\prime(g)$ is the unique Haar measure on $SO(3)$. $D_g$ is dense in 
$L^2\left(\frac{SO(3)}{SO(2)}\right)$ and it represents a
common invariant domain for the generators of the one parameter subgroups of 
$SO(3)$ i.e. $T(X_i)$ for each $i$ (see theorem 1 ch.11 $\S$1 of \cite{Group}).
Furthermore $T(\Delta)$ is the Nelson operator which, being an elliptic element
in the (right invariant) enveloping algebra of $SO(3)$, is essentially
self-adjoint in $D_G$ which, thus, represents a common invariant domain for both
$T(\Delta)$ and $T(X_i)$ for all $X_i\in\mathfrak{so}(3)$. 

Let us now refer to $\mcl\left(L^2\left(\frac{SO(3)}{SO(2)}\right)\right)$ as the set of all linear operators on
$L^2\left(\frac{SO(3)}{SO(2)}\right)$ and let $\left|\mcl\left(L^2\left(\frac{SO(3)}{SO(2)}\right)\right)\right|$ 
be the free abelian semigroup generated by the elements $\left|A\right|$ where $\left|A\right|$
is the ``absolute value''\footnote{We refer to definition of \cite{Group} of
absolute value of an operator. Pick any triple $A,B,C\in\mcl\left(L^2\left(\frac{SO(3)}{SO(2)}\right)\right)$. If 
for all $\psi\in L^2\left(\frac{SO(3)}{SO(2)}\right)$, $||C\psi||\leq ||A\psi||+||B\psi||$, then we
represent it as $|C|\leq |A|+|B|$. Thus the absolute value of an operator $A\in
\mcl\left(L^2\left(\frac{SO(3)}{SO(2)}\right)\right)$ is just a symbolic representation for the set consisting of $A$
alone.} of the operator $A$. Consequently any element
$C\in\left|\mcl\left(L^2\left(\frac{SO(3)}{SO(2)}\right)\right)\right|$ can be written as the formal sum
$C=\left|\alpha_1\right|+...+\left|\alpha_k\right|$. In particular
let us now fix $C=\sum\limits_{i=1}^3\left|T(X_i)\right|$ and 
$B=\left|T(\Delta)\right|$. According to lemma 7 in  ch.11 $\S$3 of
\cite{Group}, it exists a constant $k\leq\infty$ such that,
being $I$ the identity operator, 
$$C\leq\sqrt{\frac{3}{2}}\left|T(\Delta)-I\right|\leq\sqrt{\frac{3}{2}}\left[\;\left|
T(\Delta)\right|+\left|I\right|\;\right]$$ 
and, calling
$\left[ad (C)\right]^nB=\sum\limits_{ i_1\leq...\leq i_3}\left|T\left[ad(X_{i_3})...
ad(X_{i_1})\Delta\right]\right|$ for any $n\in\bN$ and for any $i_k=1,..,3$, it also holds that
$$\left[ad (C)\left|T(\Delta)-I\right|\;\right]^n\leq k^n\left|T(\Delta)-I\right|\leq
k^n\left[\;\left|T(\Delta)\right|+\left|I\right|\;\right].\quad\forall n\in\bN$$
Thus, $\left|T(\Delta)\right|+\left|I\right|$ analytically dominates
$\sum\limits_{i=1}^3\left|T(X_i)\right|$. The hypotheses of lemma 5 in ch.11
$\S$1 of \cite{Group} are satisfied and we may conclude that,
referring to $\overline{T(\Delta)}$ and $\overline{T(X_i)}$ as the closure
respectively of $T(\Delta)$ and $T(X_i)$ (for all $i$) on
$L^2(\bS^2)$, the domain of
$\overline{T(\Delta)}^n$ is contained in the domain of $\overline{T(X_{i_1})}...
\overline{T(X_{i_n})}$ - say $D(X_{i_1}...X_{i_n})$ - for any positive integer 
$n$ and for any finite sequence $i_1,...,i_n$. Let now consider the intersection of all the domains 
$\overline{T(\Delta)}^n$ which, up to the identification of 
$\left(\frac{SO(3)}{SO(2)}\right)$ with $\bS^2$ coincides with $\mce$. Then any 
vector in $\mce$ also lies in $D(X_{i_1}...X_{i_n})$ and in particular in the
domain of $\overline{T(X)}$ for any $X\in\mathfrak{so}(3)$. By Stone theorem
this is the set of elements in $L^2(\bS^2)$ for which the limit in (\ref{limit})
exists. Therefore if $\psi\in\mce$, $T(g)\psi$ admits all partial derivatives in
$g=e$. Let us now introduce the adjoint representation of $SO(3)$ in its Lie
algebra which maps $X\in\mathfrak{so}(3)$ into $Ad(g)X=g^{-1}Xg$ for any $g\in
\left(\frac{SO(3)}{SO(2)}\right)$. Hence a straightforward application of the properties of a generic
representation shows that $\overline{T(X)}T(g)\psi=T(g)\overline{T(Y)}\psi$
where $Y=Ad(g^{-1})X$ for any $g\in SO(3)$. This relation translates in the
chain rule $\overline{T(X_1)}...\overline{T(X_k)}T(g)\psi=T(g)\overline{T(Y_1)} 
...\overline{T(Y_k)}\psi$ where $Y_i=Ad(g^{-1})X_i$. Thus we may conclude that for all
$\psi\in\mce$
$T(g)\psi$ admits partial derivatives to all order for all $g\in SO(3)$ and 
thus, acting $T$ as the right multiplication and being
the $SO(3)$ action transitive on
$\frac{SO(3)}{SO(2)}$, $\psi$ is an infinitely
differentiable function over $\frac{SO(3)}{SO(2)}$. As
a consequence $\mce$ is contained in the space of 
infinitely differentiable functions on $\bS^2$. 

We have shown that $C^\infty(\bS^2)$ is a nuclear space and, according to
definition \ref{nuclear}, it is also a nuclear Lie group. $\Box$\\

\noindent A few remarks are in due course:\\

\remark{\label{triplet}
To each $\mathcal{E}_p$ it is possible to associate the topological dual space
$\mathcal{E}^\prime_p$ i.e. the set of continuous linear functional from
$\mce_p$ to $\bR$. It can be closed to Hilbert space with respect to the norm
$\left|\left|,\right|\right|_{-p}$ such that
$\left|\left|\psi\right|\right|_{-p}=\left|\left|A^{-p}\psi\right|\right|$ being
$\left|\left|,\right|\right|$ the $L^2(\bS^2)$-norm. We may now define
$\mathcal{E}^\prime=\bigcup\limits_p\mathcal{E}_p$ which is the topological dual
space of $\mathcal{E}$ and thus we will also refer to it
as the space of real distributions on $\bS^2$. Consequently we end up
with the following {\bf Gelfand triplet}
\beq\label{triplet2}
\mathcal{E}\subset L^2(\bS^2)\subset\mathcal{E}^\prime
\eeq
together with the set of continuous
inclusions $\mathcal{E}\hookrightarrow\mathcal{E}_p\hookrightarrow
L^2(\bS^2)\hookrightarrow\mathcal{E}^\prime_p\hookrightarrow\mathcal{E}^\prime$.
We shall denote with $\left(,\right)$ the natural pairing between $\mce^\prime$ and
$C^\infty(\bS^2)$ and it will be subject to the compatibility condition:
\beq\label{compat}
\left(\alpha(\z,\bz),\alpha^\prime(\z,\bz)\right)=\langle\alpha(\z,\bz),
\alpha^\prime(\z,\bz)\rangle_{L^2},
\eeq
for any $\alpha(\z,\bz)\in L^2(\bS^2)$ and any $\alpha^\prime(\z,\bz)\in 
C^\infty(\bS^2)$. In (\ref{compat}) $\langle,\rangle$ stands for the internal 
product in $L^2(\bS^2)$.
}\\
\remark{
The realization of the set of supertranslations as a nuclear space embedded in a
Gelfand triplet is completely different from the construction in \cite{Mc1} also
followed in \cite{DMP}.
Although the result is ultimately the same, we have decided to perform a 
different demonstration for the nuclearity of $C^\infty(\bS^2)$ since it will 
allow us a rigorous construction of the BMS invariant phase space as discussed in the
next section. On the opposite McCarthy argument in \cite{Mc1} for nuclearity of
the space of supertranslations is a straightforward extension of the
demonstration in \cite{Gelfand2} that $C^\infty\left([a,b]\right)$ with
$[a,b]\subset\mathbb{R}$ is a  nuclear space. This is a more common and 
convenient perspective whether one wants to develop the theory of 
unitary and irreducible representations for the BMS group by means of Mackey 
theory of induction.  
}\\
\remark{
If we adopt the standard topology for $SO(3,1)\sp\uparrow$ and the product
topology for $\gbms$, then one can straightforwardly conclude that the
conditions of definition \ref{nuclear} are met, thus the whole BMS group becomes 
a nuclear Lie group.
}\\

\noindent To conclude the section we wish to prove a last theorem concerning the abelian
ideal of the $\gbms$ group which will be exploited in the discussion of the
covariant wave function.\\

\teorema{\label{decompos}
Referring to $T^4$ as the closed subspace of $C^\infty(\bS^2)$ out of the real
linear combinations of the first four real spherical harmonics
$Y_{lm}(\z,\bz)$ (with $l=0,1$ and $m=-l,..,l$) and to $ST$ as the
closed\footnote{The closure is here defined for $ST$, as well as for $T^4$, with 
respect to the induced topology $\tau_p$ for $C^\infty(\bS^2)$.}
subspace of $C^\infty(\bS^2)$ out of the linear combinations of the real spherical
harmonics $\left\{Y_{lm}(\z,\bz)\right\}_{l>1}$, the following holds:
$$C^\infty(\bS^2)=T^4\oplus ST,$$
where $\oplus$ stands for the direct sum.}\\

\emph{Proof.}
The statement of the theorem can be straightforwardly proved in several
different ways if we refer to $L^2(\bS^2)$. One of the simplest consists of
recognizing that the spherical harmonics are an orthonormal complete system of
$L^2(\bS^2$) constructed according to standard harmonic functions techniques once $\bS^2$ is
identified as in the proof of theorem \ref{main1} with the symmetric space
$\frac{SO(3)}{SO(2)}$ (see chapter 10 $\S$3 in
\cite{Group}).
Thus we may claim that $L^2(\bS^2)=T^4\oplus ST$ and that, since
$C^\infty(\bS^2)\subset L^2(\bS^2)$, any $\alpha(\z,\bz)\in C^\infty(\bS^2)$ can
be univocally decomposed as
$\sum\limits_{l=0}^\infty\sum\limits_{m=-l}^l\alpha_{lm}Y_{lm}(\z,\bz)$
which converges to $\alpha(\z,\bz)$ with respect to the topology of $L^2(\bS^2)$. Furthermore take into account
that, per construction, each $Y_{lm}(\z,\bz)\in C^\infty(\bS^2)$. 

We now show that the same sum converges in the topology of $\mce\equiv
C^\infty(\bS^2)$ as constructed in theorem \ref{main1}.
Let us thus choose $\epsilon>0$ such that, for any $n\in\bN$ greater than a fixed
natural number $\bar{n}$,
$\left|\left|\alpha(\z,\bz)-\sum\limits_{l=0}^n\sum\limits_{m=-l}^l\alpha_{lm}
Y_{lm}(\z,\bz)\right|\right|<\epsilon$,
being $\left|\left|,\right|\right|$ the $L^2$-norm. Let us now consider the operator
$A=L^2+kI$ $(k>1)$ and let us evaluate 
$$\left|\left|A\left(\alpha(\z,\bz)-\sum\limits_{l=0}^n\sum\limits_{m=-l}^l\alpha_{lm}
Y_{lm}(\z,\bz)\right)\right|\right|\leq$$
$$\leq\left|\left|L^2\left(\alpha(\z,\bz)-\sum
\limits_{l=0}^n\sum\limits_{m=-l}^l\alpha_{lm}Y_{lm}(\z,\bz)\right)\right|\right|+
k\epsilon.$$
Per linearity of $L^2$ we know that 
$$L^2\left(\alpha(\z,\bz)-\sum\limits_{l=0}^n\sum\limits_{m=-l}^l\alpha_{lm}
Y_{lm}(\z,\bz)\right)=L^2\alpha(\z,\bz)+\sum\limits_{l=0}^n\sum\limits_{m=-l}^l
l(l+1)\alpha_{lm}Y_{lm}(\z,\bz).$$
We can now exploit again the harmonic function theory according to which if $\alpha^\prime(\z,\bz)\in
L^2\left(\frac{SO(3)}{SO(2)}\right)$ then the sum
\begin{gather*}
\alpha^\prime(\z,\bz)=\sum\limits_{l=0}^\infty\alpha^\prime_{lm}Y_{lm}(\z,\bz),\\
\alpha^\prime_{lm}=\int_{\frac{SO(3)}{SO(2)}}d\mu(\z,\bz)\alpha^\prime(\z,\bz)
Y_{lm}(\z,\bz)
\end{gather*}
converges in the topology of $L^2\left(\frac{SO(3)}{SO(2)}\right)$ and the decomposition is unique.

Furthermore, since
$$\left[L^2\alpha\right](\z,\bz)=\sum\limits_{l=0}^\infty\alpha^\prime_{lm}Y_{lm}
(\z,\bz),$$
we may conclude, by means of the decomposition of
$\alpha(\z,\bz)$ in spherical harmonics, that
$$\alpha^\prime_{lm}=l(l+1)\alpha_{lm}.$$
Consequently 
$$\left|\left|L^2\left(\alpha(\z,\bz)-\sum\limits_{l=0}^n\sum\limits_{m=-l}^l
\alpha_{lm}Y_{lm}(\z,\bz)\right)\right|\right|=$$
$$=\left|\left|[L^2\alpha(\z,\bz
)]-\sum\limits_{l=0}^n\sum\limits_{m=-l}^l l(l+1)\alpha_{lm}Y_{lm}(\z,\bz)
\right|\right|\leq\epsilon,$$
for sufficiently large $n$; it descends
\begin{equation*}
\left|\left|A\left(\alpha(\z,\bz)-\sum\limits_{l=0}^n\sum\limits_{m=-l}^l\alpha_{lm}
Y_{lm}(\z,\bz)\right)\right|\right|\leq(k+1)\epsilon,
\end{equation*}
i.e. the sum $\sum\limits_{l=0}^n\sum\limits_{m=-l}^l\left[l(l+1)+k\right]
\alpha_{lm}Y_{lm}(\z,\bz)$ converges to $A\alpha(\z,\bz)$ in the topology of 
$L^2(\bS^2)$.\\
The same reasoning leads to the same conclusion with respect to 
\begin{equation*}
A^p\left[\alpha(\z,\bz)-\sum\limits_{l=0}^n\sum\limits_{m=-l}^l\left[l(l+1)+k\right]
\alpha_{lm}Y_{lm}(\z,\bz)\right]
\end{equation*}
for any integer $p$. Consequently the series 
$\sum\limits_{l=0}^n\sum\limits_{m=-l}^l\alpha_{lm}Y_{lm}(\z,\bz)$ converges to
$\alpha(\z,\bz)$ with respect to $\left|\left|,\right|\right|_p$ in each $\mce_p$ as 
introduced in theorem \ref{main1}. Consequently, per definition, the series 
converges as well to $\alpha(\z,\bz)$ in $\mce\equiv C^\infty(\bS^2)$ with 
respect to the induced topology $\tau_p$. 

The uniqueness of the decomposition is accordingly traded from $L^2(\bS^2)$ to
$C^\infty(\bS^2)$ and the hypotheses of closure in $C^\infty(\bS^2)$ for $ST$ is
justified. $\Box$\\

\remark{
It is interesting to notice that $\left(C^\infty(\bS^2),\tau_p\right)$
constructed as in theorem \ref{main1} can also be interpreted as a \emph{strong
inverse limit of (abelian) Hilbert} Lie (ILH) groups as discussed in \cite{Omori}. 
Furthermore, bearing in mind theorem \ref{decompos}, the decomposition 
$C^\infty(\bS^2)=T^4\oplus ST$ is a \emph{ILH-splitting} and both $T^4$ and $ST$
are ILH-subgroups of $C^\infty(\bS^2)$. These considerations are automatically
traded to the full $\gbms$ group endowed with the product topology and they will be exploited 
in the forthcoming discussions.
}

\section{BMS free field theory}\label{tre}
The development of a field theory invariant under a BMS transformation has been already discussed 
in previous papers. Nonetheless we shall recast some of the already known results either for sake of
completeness either since they play a pivotal role in the next two sections. Thus,
as a starting point, we need to review some of the concepts and of the nomenclatures
of \cite{Arcioni} and, mainly, of \cite{DMP} {\it i.e.} we will devout the section 
to sketch the construction of the BMS unitary and irreducible
representations (irreps.) and consequently of the induced wave functions. 

\subsection{BMS unitary and irreducible representations} As we have demonstrated
in the previous section, the $\gbms$ group is an infinite dimensional nuclear
Lie group with a semidirect product structure. Thus, in order to develop the theory of
irreps. for such a group we shall make use of the inductions techniques as developed by Mackey (see in 
particular \cite{Simms,Mackey} and the recent review in \cite{LLedo})
and extended to a semidirect product with an infinite dimensional abelian ideal by Piard
in \cite{Piard}.

As a first step it is much more
convenient to replace $SO(3,1)\sp\uparrow$, the proper ortochronous subgroup of 
the Lorentz group, 
with its universal cover $SL(2,\bC)$. At a level of theory of representations
such operation is not pernicious since, beyond those of $\gbms$, it introduces only a further irreducible
representation induced from the $\bZ_2$ subgroup. Thus
from now on we will switch from $\gbms$ to $\tgbms=SL(2,\bC)\ltimes
C^\infty(\bS^2)$ which is still a nuclear Lie group.

The second step in Mackey machinery consists on constructing a ``character''  by means
of the following proposition, proved in section 3.2 of
\cite{DMP}:\\

\proposizione{
Given an abelian topological group $A$, a {\bf character} is a continuous group
homomorphism $\chi:A\to U(1)$, the latter being equipped with the natural
topology induced by $\bC$. If $A=\mathcal{E}\equiv C^\infty(\bS^2)$ then it
exists a unique real distribution $\beta\in\mathcal{E}^\prime$ such that
$\chi\left(\alpha(\z,\bz)\right)=\exp[i\left(\beta,\alpha(\z,\bz)\right)]$ for any 
$\alpha(\z,\bz)\in C^\infty(\bS^2)$. Here 
$\left(\beta,\alpha(\z,\bz)\right)$ stands for the natural dual pairing\footnote{Since
$\mathcal{E}^\prime$ is a space of distributions we will also refer to the
pairing $(\beta,\alpha)$ as the evaluation of the distribution $\beta$ on the
test function $\alpha$.} between $C^\infty(\bS^2)$
and its topological dual $\mathcal{E}^\prime$ constructed in remark 
\ref{triplet}.}\label{chara}\\

\remark{
The set of characters $\overline{A}$, equipped with the product operation 
$$\left(\chi_1\chi_2\right)\left(\alpha(\z,\bz)\right)=\chi_1
\left(\alpha(\z,\bz)\right)\chi_2\left(\alpha(\z,\bz)\right)\quad\forall
\alpha\in\mce$$ 
is an abelian group called the {\bf dual character group}.
}\\
The third step in Mackey's machinery, applied to a regular semidirect product,
consists of the identification of three key structures:\\

\definizione{\label{orbit}
Consider $G=B\ltimes A$ as the regular semidirect product between a topological abelian
group $A$ and any group $B$. Then for any $\chi\in\overline{A}$, we may
associate:
\begin{itemize}
\item the {\bf orbit} $\mco_\chi\subset\overline{A}$ as the set 
$$\mco_\chi=\left\{\chi^\prime\in\overline{A}\;\left|\right.\;\exists g\in
G\;with\;\chi^\prime=g\chi\right\},$$
where $g\chi(a)=\chi(g^{-1}a)$ for any $a\in A$ and for any $g\in G$.
\item the {\bf isotropy group} $H_\chi\doteq\left\{g\in G\;\left|\right.\;g\chi=\chi\right\}$.
\item the {\bf little group} $L_\chi\subset H_\chi$ as the subset $\left\{g\in H_\chi\;\left|\right.\; g=(\Lambda,0)
\in G\right\}$. 
\end{itemize}
} 

\remark{
Referring to $\tgbms$ the construction of the structures outlined in definition
\ref{orbit} is rather simple since it only requires the identification of the little groups. 
As shown in \cite{DMP} and in
\cite{Mc1}, given a fixed character $\chi$, its isotropy group is $H_\chi=
L_\chi\ltimes C^\infty(\bS^2)$ whereas the associated orbit $\mco_\chi$ is the
quotient $\frac{\tgbms}{H_\chi}\sim\frac{SL(2,\bC)}{L_\chi}$. Furthermore it
turns out that all possible little groups $L_\chi$ are closed subgroups of $SL(2,\bC)$ 
namely $SU(2)$, $SO(2)$, $\Delta$ the double cover of the two dimensional Euclidean subgroup,
$SL(2,\bR)$ and the set of all cyclic, alternating and dihedral finite
dimensional groups of order $n\geq 2$.
}\\
   
\remark{\label{critical}
According to the above definition the orbit $\mco_\chi$ should be thought as
embedded in the space of characters and as generated by the action of
$\frac{SL(2,\bC)}{L_\chi}$ on $\chi$ where $L_\chi$ satisfies $L_\chi\chi=\chi$.
Nonetheless an equivalent point of view
arises exploiting proposition \ref{chara} according to which
$\chi(\alpha)=e^{i\left(\beta,\alpha(\z,\bz)\right)}$ for a unique choice of
$\beta\in\mce^\prime$. Thus, since
$\Lambda\chi(\alpha(\z,\bz))=\chi(\Lambda^{-1}\alpha(\z,\bz))=e^{i\left(\Lambda
\beta,\alpha(\z,\bz)\right)}$ for any $\Lambda\in SL(2,\bC)$, the defining 
equation for $L_\chi$, i.e. $L_\chi\chi=\chi$, can be traded with an analogous 
equation in $\mce^\prime$ i.e. $L_\chi\bar{\beta}=\bar{\beta}$. Consequently also the orbit
$\mco_\chi$ in the character space is canonically isomorphic to the orbit
embedded in $\mce^\prime$ and generated by the action of $\frac{SL(2,\bC)}{L_\chi}$ 
on $\bar{\beta}$. For this reason, from now on, we will stick to the much more
convenient perspective $\mco_\chi\hookrightarrow\mce^\prime$ though we will
retain the pedex $\chi$ for later convenience.
}\\

Let us now still focus our attention specifically to $\tgbms$ and let us switch to the more convenient
language of fiber bundles. We introduce the
\emph{Mackey bundle} $G_\chi=\tgbms\left[\mco_\chi, H_\chi,\tau\right]$ 
with $\tgbms$ as total space, the orbit of $\chi$ as base space, the associated isotropy group 
$H_\chi$ as typical fiber whereas the projection $\tau:G\to\mco_\chi$ is suitably chosen case by
case\footnote{The existence of $\tau$ is not a priori granted in a general 
scenario but, in the BMS setting, such projection maps have been explicitly 
identified for all possible little groups \cite{Arcioni}.}. 

Furthermore, bearing in mind that $\mco_\chi$ is $L_\chi\ltimes C^\infty(\bS^2)$, we may 
select a unitary and irreducible representation $\Sigma$ of $\mco_\chi$ acting on a
suitably chosen Hilbert space $\mch$. Moreover, for any but fixed $\Sigma$ and 
for any $g=(\Lambda,\alpha)\in H_\chi$, we may rewrite $\Sigma(g)$ as $\chi(
\alpha)\sigma(g)$ where $\sigma$ is a unitary irrep. of $L_\chi$.

We may proceed constructing the \emph{associated
Hilbert bundle} to $G_\chi$ as $\mathbb{H}=G_\chi\times_\Sigma\mch$ which is a bundle 
topologically equivalent to the Cartesian product between $G_\chi$ and $\mch$
whose elements are equivalence classes 
\begin{eqnarray*}
[g,\psi]=&\negmedspace\left\{(g,\psi)\in G_\chi\times\mch\;\left|\right. (g,\psi)\sim 
(g^\prime,\psi^\prime)\quad\right.\\
& \left.\quad\; iff \quad\exists \tilde g\in
\mco_\chi\;\left|\right.\; g^\prime=\tilde g g,\;\wedge\;
\psi^\prime=\Sigma(\tilde g)\psi\right\}.
\end{eqnarray*}  
$\mbH$ can be interpreted as a bundle with $\mco_\chi$ as the base space,
$\mch$ as the typical fiber whereas the projection $\tilde\tau:\mbH\to\mco_\chi$ maps
$[g,\psi]$ in $\tilde\tau\left([g,\psi]\right)\doteq\tau(g)$.

We are now in position to 
apply the standard induction technique in order to define a unitary and
irreducible representation of the full $\tgbms$-group. Let us thus start introducing the
set of smooth sections of the associated Hilbert bundle $\mbH$ which, up to the
choice of a global Borel section $s:\mco_\chi\to\tgbms$ for the Mackey bundle $G_\chi$, can be
characterized as the set:
\beq
\Gamma(\mbH)_s=\left\{\Phi_s:\mco_\chi\longrightarrow\mch\;\left|\right.\;\Phi_s\in
C^\infty(\mco_\chi,\mch)\right\}.
\eeq 
We may now exploit theorem 3.1 in \cite{DMP} in order to associate to $\mco_\chi$
its unique quasi-invariant measure class $\left[\mu\right]$. Thus we may close the
space $\Gamma(\mbH)_s$ to an Hilbert space as 
\beq\label{sec1} 
\widetilde\mch_{s,\mu}=\left\{\Phi_s\in C^\infty(\mco_\chi,\mch)\;\left|\right.\;
\int\limits_{\mco_\chi}d\mu(p)\langle\Phi(p),\Phi(p)\rangle <\infty\right\},
\eeq
where $p\in\mco_\chi$, $\mu$ is any representative of $\left[\mu\right]$ and $\langle,\rangle$
is the scalar product in $\mch$. 

Furthermore each element in $\widetilde\mch_{s,\mu}$ inherits the natural $\tgbms$-action
\beq\label{induced}
\left(g\Phi_s\right)(p)=\sqrt{\frac{d\mu(g^{-1}p)}{d\mu (p)}}\left(g\Phi_s\right)
(g^{-1}p),\;\;\forall g=(\Lambda,\alpha)\in SL(2,\bC)\ltimes C^\infty(\bS^2)
\eeq
which can be rewritten in the more common and convenient form \cite{Simms}:
\begin{gather}
(\Lambda\Phi)(p)=\sqrt{\frac{d\mu(\Lambda^{-1}p)}{d\mu
(p)}}\sigma\left(s(p)^{-1}\Lambda s(\Lambda^{-1} p)\right)
\Phi(\Lambda^{-1}p),\\
(\alpha\Phi)(p)=\chi(\alpha)\Phi(p),
\end{gather}
where $\frac{d\mu(\Lambda^{-1}p)}{d\mu (p)}$ is the Radon-Nikodym derivative and
where, bearing in mind remark \ref{critical}, $\chi(\alpha)=e^{i(p,\alpha)}$. 
Furthermore the following holds:
\begin{itemize}
\item[$\bullet$] applying lemma 1 in $\S$16 of \cite{Group}, we may conclude that $\widetilde\mch_{s,\mu}$, 
whose elements satisfy (\ref{induced}), is isomorphic to the Hilbert space $L^2(\mco_\chi,\mu)\otimes\mch$.
\item[$\bullet$] applying Mackey's theorem (see \cite{Mackey} or chapter 16 in
\cite{Group}) (\ref{induced}) is a strongly continuous unitary  
representation of the $\tgbms$ group induced from $\Sigma=\chi\sigma$.
\item[$\bullet$] all the induced $\tgbms$ unitary representations are
irreducibles. Nonetheless a complete list is not available at the moment since 
all the irreps. must arise either from a transitive
$SL(2,\bC)$-action on $\mce^\prime$ or from a cylinder measure with respect to
which the $SL(2,\bC)$-action is strictly ergodic. The latter condition is rather
difficult to deal with and the problem of studying it in detail has not been
addressed yet.  
\end{itemize}
We may summarize the information from the above discussion in the following 
statement:\\

\definizione{\label{induc}
We call $\tgbms$ {\bf induced wave function} (or $\tgbms$ free field) any map in (\ref{sec1}) which satisfies 
(\ref{induced}) i.e. it is a square integrable function over $\mco_\chi$ with values in a suitably chosen target Hilbert space 
$\mch$ and it transforms under a unitary and irreducible induced representation 
of the $\tgbms$-group.
}\\

\noindent In order to complete the analysis of free fields exploiting inducing techniques
it is also necessary to construct the full set of Casimir invariants for the
unitary $\tgbms$-representations. Bearing in mind the example of the Poincar\'e
group, one hopes to give a group-theoretical definition to the notion of mass
for an induced $\tgbms$-field and to univocally characterize the orbit by the
lone value of the invariants.

In order to achieve this goal we exploit the following proposition (see also 
chapter 4 in \cite{Gelfand2}):\\
\proposizione{\label{annihil}
Given any subspace $V$ of a locally convex linear topological space $\Psi$ any 
linear continuous functional $\beta:V\to\mathbb{C}$ can be extended to a
functional on all $\Psi^\prime$. Furthermore if we introduce the annihilator of
$V$ as:
\beq
V^0=\left\{\beta\in\mce^\prime\;\left|\right.\;\left(\beta,v\right)=0,\forall
v\in V\right\},
\eeq
the following holds
\begin{enumerate}
\item the factor space $\frac{\Psi^\prime}{V^0}$ is the dual space of $V$
\item If $V$ is a $d$-dimensional subspace of $\Psi$ (with $d<\infty$), then
also $\frac{\Psi^\prime}{V^0}$ is $d$-dimensional
\end{enumerate}}

\emph{Proof.}
Let us take any continuous linear functional $\beta\in V^\prime$; continuity implies that it
exists a neighborhood $U$ of $0\in\Psi$ such that
$\left|\left(\beta,\alpha\right)\right|\leq 1$ for all $\alpha\in U\cap V$. Let
us choose now any absolutely convex neighborhood $U'\subset U$ whose existence
is granted by the local convexity of $\Psi$. We now consider $U'$ as the unit
sphere in $\Psi$ of a seminorm such that
$\left|\left|\alpha\right|\right|=\left[sup\left(\lambda\right)\right]^{-1}$ where
$\lambda\alpha\in U'$ for all $\alpha\in\Psi$. Per construction we end up with
$\left|\left(\beta,\alpha\right)\right|\leq\left|\left|\alpha\right|\right|$ for any
$\alpha\in V$. Due to Hahn-Banach theorem (straightforwardly adapted to a space
with seminorms), the functional $\beta$ admits an extension $\bar\beta$ on all 
$\Psi$ which furthermore is bounded i.e.
$\left|\left(\bar\beta,\alpha\right)\right|\leq\left|\left|\alpha\right|\right|$ for all
$\alpha\in\Psi$. Thus it follows that
$\left|\left(\bar\beta,\alpha\right)\right|\leq 1$ for all $\alpha\in U'$ i.e.
$\bar\beta$ is also continuous relatively to the topology on $\Psi$. This
concludes the first part of the demonstration.

Let us now consider any $\beta\in\Psi^\prime$ which is also a functional on $V$
being $V$ a subspace of $\Psi$. Two functionals $\beta_1$, $\beta_2$ do coincide
on $V$ iff they belong to the same coset in $\frac{\Psi^\prime}{V^0}$. Clearly
to any $[\beta]\in\frac{\Psi^\prime}{V^0}$ corresponds an element on
$V^\prime$ and, if $[\beta_1]\neq[\beta_2]$ on $\frac{\Psi^\prime}{V^0}$, then
the corresponding functionals on $V^\prime$ are distinct. The point consists
of showing that every linear functional on $V^\prime$ can be constructed in the
following way. Let $\beta_0$ be any functional in $V^\prime$. Then, by
Hahn-Banach theorem, it can be extended to a linear functional on $\Psi^\prime$
and all the possible extensions coincide on $V$ i.e. they belong to the same
coset relatively to $V^0$. Consequently every linear functional on $V$
corresponds to an element of the factor space $\frac{\Psi^\prime}{V^0}$. 

To conclude suppose now that $V$ is finite $d$-dimensional. Then the above
result immediately implies that also $V^\prime$ and consequently 
$\frac{\Psi^\prime}{V^0}$ is $d$-dimensional. $\Box$\\

In the $\tgbms$ setting, this proposition can be exploited considering the
subspace consisting of the real linear combinations of the first four real
spherical harmonics $Y_{lm}(\z,\bz)\subset C^\infty(\bS^2)$ with $l=0,1$ and
$m=-l,...,l$. This is a four-dimensional subspace which we will refer to as
$T^4$ and which, furthermore, is
invariant under the $SL(2,\bC)$ action induced by (\ref{product}). Thus, since,
according to theorem \ref{nuclear}, $C^\infty(\bS^2)$ is a nuclear space and
thus a locally convex linear topological space, we can introduce the projection
\beq\label{proj}
\pi:\mce^\prime\longrightarrow\frac{\mce^\prime}{(T^4)^0}\sim\left(T^4\right)^\prime, 
\eeq
where the isomorphism between $\left(T^4\right)^\prime$ and 
$\frac{\mce^\prime}{(T^4)^0}$ is $SL(2,\bC)$ invariant.
The map (\ref{proj}) enjoys the following remarkable properties whose 
demonstration is given in \cite{DMP,Mc1} (though with slightly different
techniques and nomenclatures)\\

\proposizione{\label{mass2}
Let $\beta\in\mce^\prime$ and let $\left\{Y^*_{lm}\right\}$ (with $l=0,1$ and
$m=-l,...,l$) be the base of $\left(T^4\right)^\prime$ constructed in such a way that
$\left(Y^*_{lm},Y_{l'm'}\right)=\delta_{ll'}\delta_{mm'}$ where $(,)$ refers to
the natural pairing between $C^\infty(\bS^2)$ and $\mce^\prime$. Consider
$$\pi(\beta)=\sum\limits_{l=0}^1\sum\limits_{m=-l}^l a_{lm}Y^*_{lm},$$
from which we can extract the four vector\footnote{The extraction of a 4-vector
from the coefficients of the expansion in dual spherical harmonics a posteriori
justifies the symbol $T^4$ for the subspace of $C^\infty(\bS^2)$ generated by
$\left\{Y_{lm}(\z,\bz)\right\}$ with $l=0,1$ and $m=-l,...,l$.}
$$\widehat{\pi(\beta)}_\mu=-\sqrt{\frac{3}{4\pi}}(a_{00},a_{1-1},a_{10},a_{11}).$$
Moreover, if one defines the real bilinear form $B$ on $\mce^\prime$ such that
\beq\label{mass}
B(\beta_1,\beta_2)=\eta^{\mu\nu}\widehat{\pi(\beta_1)}_\mu\widehat{\pi(\beta_2)}
_\nu,\quad\forall\beta_1,\beta_2\in\mce^\prime
\eeq
then $B$ turns out to be $SL(2,\bC)$ invariant and a Casimir invariant for the
$\tgbms$ unitary and irreducible representations.} \\
\remark{
In analogy with the Poincar\'e counterpart, we will refer to (\ref{mass}) as the
defining relation for the $\tgbms$ squared mass $m^2$. Furthermore such
proposition justifies a posteriori the reason for the name of
\emph{space of supermomenta} for $\mce^\prime$ which is common in the physical
literature.
}\\
\remark{
In \cite{Mc1} McCarthy showed that the values of $m^2$ together with the sign of
$\widehat{\pi(\beta)}_0$ univocally characterize the orbits only for the little
group $SU(2)$. In all other cases to each orbit it is possible to assign a
single value of $m^2$ which does not completely identify/describe it; furthermore there
is only one connected subgroup of $SL(2,\bC)$ which admits $m^2=0$ namely
$\Delta$ the double cover of the two dimensional Euclidean subgroup: exactly the same
little group associated to the massless fields in a Poincar\'e invariant theory on
Minkowski spacetime.
}\\

\noindent Though we have fully characterized the full set of $\tgbms$ induced free fields,
we need to remember that ultimately our goal is to develop a Lagrangian and an Hamiltonian
formulation of $\tgbms$ free field theory. Thus it would be rather prohibitive to deal
contemporary with all the possible cases outlined above and we shall make use of
a simple but exhaustive ``working'' example namely the $\tgbms$ scalar field. We
distinguish between two cases \cite{DMP,Mc1}:
\begin{enumerate}
\item {\bf the $\tgbms$ real massive scalar field} which is a map $\Phi\in
L^2(\frac{SL(2,\mathbb{C})}{SU(2)_\chi},\mu)$ whose orbit is generated by the
action of $\frac{SL(2,\bC)}{SU(2)}$ on the real distribution
$\bar\beta=\sqrt{\frac{4\pi}{3}}mY^*_{00}$. Furthermore we stress that, since
$\bar\beta\in \left(T^4\right)^\prime$, we can exploit the $SL(2,\bC)$ invariant
isomorphism on the right hand side of (\ref{proj}) to conclude that the whole orbit is contained in
$\left(T^4\right)^\prime$. If we now choose
$\mu$ as the $SL(2,\bC)$-invariant measure on the hyperboloid 
$\frac{SL(2,\mathbb{C})}{SU(2)_\chi}$, then $\Phi$ transforms under a
$\tgbms$ action as 
\beq\label{massive}
\left(g\Phi\right)(\beta)=e^{i\beta(\alpha)}\Phi(\Lambda^{-1}\beta),\quad\forall
g=(\Lambda,\alpha)\in\tgbms\;\wedge\;\beta\in\frac{SL(2,\bC)}{SU(2)}\bar\beta
\eeq
\item
{\bf the $\tgbms$ real massless scalar field} which is a map $\Phi\in
L^2(\frac{SL(2,\mathbb{C})}{\Delta_\chi},\mu)$ whose orbit is generated by the
$\frac{SL(2,\bC)}{\Delta}$ action on the real distribution
$\bar\beta=\left(C\delta+K\delta^{(2,2)}+S|z|^{-6}\right)\left(1+|z|^2\right)^3$
where $\delta^{(2,2)}$ represents the derivative of the $\delta$ function twice
respect to the variable $\z$ and $\bz$ whereas $K,S\in\bR$ and $C\in\bR-\left\{0\right\}$. 
Furthermore, as in the massive case, the fixed point $\bar\beta$ lies in
$\left(T^4\right)^\prime$ and thus we may exploit (\ref{proj}) to conclude that
the whole orbit lies in $\left(T^4\right)^\prime$. If we choose
$\mu$ as an $SL(2,\bC)$-invariant measure on the light-cone
$\frac{SL(2,\mathbb{C})}{\Delta_\chi}$, then $\Phi$ transforms under a
$\tgbms$ action as 
\beq\label{massless}
\left(g\Phi\right)(\beta)=e^{i(\beta,\alpha)}\Phi(\Lambda^{-1}\beta)\quad\forall
g=(\Lambda,\alpha)\in\tgbms\;\wedge\;\beta\in\frac{SL(2,\bC)}{\Delta}\bar\beta
\eeq
Furthermore we shall now remember theorem 3.2 in \cite{DMP} according to which
only the field living on the orbit with $K=S=0$ coincide with the projection on $\Im^+$ - i. e. 
null infinity - of
a solution for the massless Klein-Gordon equation conformally coupled to gravity
in the bulk of any asymptotically flat and globally hyperbolic spacetime. For
this reason when we will refer from now on to a real $\tgbms$ massless scalar
field we will consider implicitly this physically relevant case. Nonetheless
most of results and all the techniques we will make use of may be
straightforwardly extended to the general case.
\end{enumerate}

\section{The covariant wave function and the associated functional spaces}\label{quattro} 
The aim of this section is to fill a gap in the discussion of field
theory at future null infinity as it is appeared up to now in the literature. In \cite{Arcioni} and 
\cite{DMP} the key ingredient to describe a BMS invariant theory was the so-called canonical or 
induced point of view according to which a BMS free field is a function transforming under a unitary and 
irreducible representation of the full BMS group. On the opposite the covariant perspective, though 
fully equivalent to the canonical one and more common in physics, has not been dealt with in detail. Since
this latter point of view is ultimately the most natural one to deal with a Lagrangian or 
an Hamiltonian formulation of the  BMS field theory, we need to amend such lack. 

Thus, in order to get over the concept of induced wave function as free 
field introduced in definition \ref{induc}, the starting point consists on noticing that, for a
semidirect product group, the Mackey bundle can be traded with a different one:
\beq\label{covbundle}
G^\prime=\tgbms\left[C^\infty(\bS^2),SL(2,\bC),\tau^\prime\right].
\eeq
The $\tgbms$ group is still the total space but $\mce=C^\infty(\bS^2)$ acts as a base
space whereas $SL(2,\bC)$ is the typical fiber and $\tau^\prime$ is the natural
projection mapping $g=\left(\Lambda,\alpha(\z,\bz)\right)\in\tgbms$ to $\tau(g)=
\alpha(\z,\bz)$. 

We can now exploit either theorem \ref{main1} either remark \ref{triplet} to introduce 
the regular semidirect product $SL(2,\bC)\ltimes \mce^\prime$ with the composition
rule between $g=(\Lambda,\beta)$ and $g^\prime=(\Lambda^\prime,\beta^\prime)$ as
$$g\odot g^\prime=(\Lambda\Lambda^\prime, \beta+\Lambda\beta^\prime),$$
where the $SL(2,\bC)$-action on any element of $\mce^\prime$ is 
$$\left(\Lambda\beta^\prime,\alpha(\z,\bz)\right)=\left(\beta^\prime,\Lambda^{-1}\alpha
(\z,\bz)\right),\quad\forall\alpha(\z,\bz)\in C^\infty(\bS^2)$$
being $\Lambda^{-1}\alpha(\z,\bz)=(K_{\Lambda} \circ \Lambda^{-1})\cdot (\alpha(\z,\bz)
\circ \Lambda^{-1})$ the action of $SL(2,\bC)$ on a smooth function on $\bS^2$ as in
(\ref{product}).\\
Thus we can introduce
\beq\label{covbundle2}
\widetilde{G}=\widetilde{G}\left[\mce^\prime,SL(2,\bC),\tilde\tau^\prime\right],
\eeq
which is a bundle defined as (\ref{covbundle}) merely substituting $\mce^\prime$ to 
$C^\infty(\bS^2)$. Furthermore, considering both $G^\prime$ and $\widetilde{G}$ as
principal bundles and remembering (\ref{triplet}), we can embed $G^\prime$ in
$\widetilde{G}$ by means of the natural homomorphism $i:G^\prime\to\widetilde{G}$
which maps $(\Lambda,\alpha(\z,\bz))\in G^\prime$ into the correspondant point in
$\widetilde G$.\\

\remark{
From a physical perspective the space $\mce^\prime$ is usually referred to as the space of 
supermomenta since it represents the dual of $C^\infty(\bS^2)$, the space of supertranslations.
This nomenclature, originated (see \cite{Mc1,Arcioni}) either in analogy with the Minkowski counterpart
where momenta are duals to translations either due to the identification of the $\tgbms$ group with 
$SL(2,\bC)\ltimes L^2(\bS^2)$ instead of $SL(2,\bC)\ltimes C^\infty(\bS^2)$, is rather inconvenient.  
From one side the enlargement of the abelian ideal of to an Hilbert space, though useful for calculations, 
is incorrect from the perspective of the holographic principle since, as shown in \cite{DMP}, only within a nuclear topology
such as the one associated to $C^\infty(\bS^2)$ it is possible to coherently interpret bulk data
in terms of boundary ones. From the other side, it allows to identify an isomorphism between
the supertranslations and the supermomenta applying Riesz theorem on $L^2(\bS^2)$. Such a result
does not hold in the generic nuclear scenario and, furthermore, we will show that a function with support on the space
of supermomenta - i.e. $\mce^\prime$ - cannot never be mapped into a function on the space
of supertranslations by means of a Fourier transform. This is in net contrast with the usual paradigm of 
a field theory in Minkowski spacetime and with the usual physical interpretations of translations and momenta.
} \\

The next step consists of following closely the road outlined in the previous subsection; let
us thus fix a separable Hilbert space $\mch^\prime$ and an $SL(2,\bC)$ representation
$\rho$ acting on it. Then we may construct the associated bundle to $\widetilde{G}$ as
$$\mbH^\prime=\widetilde{G}\times_{SL(2,\bC)}\mch^\prime$$
which is the set of equivalence classes of points
\begin{eqnarray*}
\left[g,\psi\right]= &\hspace{-.5cm}\left\{(g,\psi)\in\widetilde{G}\times\mch^\prime\;\left|
\right.\;(g,\psi)\sim(g^\prime,\psi^\prime)\right.\;iff\\
& \left.\exists \Lambda\in
SL(2,\bC)\;\left|\right.\;g^\prime=\Lambda g\;\wedge\;\psi^\prime=\rho(\Lambda)
\psi\right\}.
\end{eqnarray*}
A tempting conclusion would now lead to define a new set of wave function as the set
of sections for $\mbH^\prime$ endowed with a suitable regularity conditions. At this 
stage this is still not possible since, fixing the section
$s:\mce^\prime\to\widetilde{G}$ such that $\beta\mapsto s(\beta)=(e,\beta)$ being $e$
the identity element in $SL(2,\bC)$, a section of $\mbH^\prime$ is a map
$\tilde\Phi:\mce^\prime\to\mch^\prime$. Thus we need to introduce a suitable notion of
square-integrability on set of functions defined over the topological dual space of a
nuclear space. In order to achieve this goal we shall make use of the Minlos 
theorem (see \cite{Kuo} and in particular \cite{Gelfand2} for a proof):\\

\teorema[Minlos]{
Given a real nuclear space $V$ and its topological dual space $V^\prime$, the
map $\varphi:V\to\bC$ is the characteristic function of the unique probability
measure $\nu$ on $V^\prime$ such that - calling $(,)$ the pairing
between $V^\prime$ and $V$
$$\varphi(v)=\int\limits_{V^\prime}e^{i
(v^\prime,v)}d\nu(v^\prime)\quad\forall v\in V,$$
iff $\varphi(0)=1$, $\varphi$ is continuous on $V$ and positive definite i.e. for any
n-tuple of complex numbers $\left\{z_i\right\}_{i=1}^n$ and of elements in $V$,
say $\left\{v_i\right\}_{i=1}^n$
$$\sum\limits_{j,k=1}^n z_j\bar{z}_k\varphi(v_i-v_k)\geq 0.$$}\\

\begin{lemma}\label{measure}
Fixing the nuclear space $\mce=C^\infty(\bS^2)$ and its topological dual space
$\mce^\prime$ along the lines of remark \ref{triplet}, the complex valued 
function 
\beq\label{Minlos}
\varphi(\alpha(\z,\bz))=e^{-\frac{i}{2}\left|\left|\alpha(\z,\bz)\right|\right|_{L^2}}
\eeq
is the characteristic function of a unique probability measure $\nu^\prime$ on 
$\mce^\prime$.
\end{lemma}\\

\emph{Proof.}
The demonstration is similar to the standard one for the Schwartz space of
real-valued rapidly decreasing test functions on $\bR$. As a matter of fact it is
straightforward to realize that $\varphi$ is either continuous either equal to
$1$ if evaluated in $0\in C^\infty(\bS^2)$. 

We need only to verify the positivity of (\ref{Minlos}). Let us consider any 
n-tuple of complex numbers  $\left\{z_i\right\}_{i=1}^n$ and let us call with $C\subset C^\infty(\bS^2)$
the subspace (with norm $\left|\left|,\right|\right|_{L^2}$) spanned by any but 
fixed n-tuple of smooth functions over $\bS^2$, say 
$\left\{\alpha_i(\z,\bz\right\}_{i=1}^n$. Referring to the standard Gaussian measure
on $C$ with $\mu_C$, then any $\alpha(\z,\bz)\in C$ satisfies
$$\int\limits_C
d\mu_C(\alpha^\prime)e^{i\left(\alpha^\prime(\z,\bz),\alpha(\z,\bz)\right)}=e^{i
\left|\left|\alpha(\z,\bz)\right|\right|_{L^2}},$$
where $\left(,\right)$ is the internal product in $L^2(\bS^2)$. Consequently
\begin{gather}
\sum\limits_{j,k=1}^n z_j\bar{z}_k\varphi(\alpha_i(\z,\bz)-\alpha_k(\z,\bz))=
\sum\limits_{j,k=1}^n\int\limits_C d\mu_C(\alpha^\prime)
e^{i\left(\alpha^\prime(\z,\bz),\alpha_j(\z,\bz)-\alpha_k(\z,\bz)\right)}=\notag\\
=\int\limits_C d\mu_C(\alpha^\prime)\sum\limits_{j=1}^n\left|
e^{i\left(\alpha^\prime(\z,\bz),\alpha_j(\z,\bz)\right)}\right|^2\geq 0,\notag 
\end{gather}
which grants us that $\varphi$ satisfies the conditions of Minlos theorem. $\Box$\\

The pair $(\mce^\prime,\nu)$ plays in the BMS field theory the same role that
the space of momenta $(\bR^4,d^4x)$ plays for a Poincar\'e invariant field 
theory over Minkowski spacetime $\bM^4$. It it thus natural to ask ourselves if
we can define a natural counterpart in the BMS setting also for
$L^2(\bR^4,d^4x)$ as well for $\mathcal{S}(\bR^4)$, the set of rapidly
decreasing test functions over $\bR^4$ and the space of tempered distributions
$\mathcal{S}^\prime(\bR^4)$. In order to deal with this question which is
fundamental in order to define a covariant BMS (free and interacting) field 
theory, we still resort to the powerful techniques of white noise distribution
theory \cite{Kuo,Hida}. \\

\definizione{
We call the \emph{space of square-integrable functions over the supermomenta}
the set of equivalence classes of maps 
\beq
L^2(\mce^\prime,\tilde\mch,\nu)=\left\{\psi:\mce^\prime\to\tilde\mch\;\left|
\right.\;\int\limits_{
\mce^\prime}\langle\psi(\beta),\psi(\beta)\rangle d\nu(\beta)<\infty\right\},
\eeq
where $\langle,\rangle$ is the internal product on $\tilde\mch$. Two functions
are equivalent if they agree everywhere except in a set of zero measure.
This space is also referred\footnote{In \cite{Kuo,Hida2} this space is also
called ``white noise space'' though $\mce^\prime$, the space of real distributions
over $\bS^2$ is traded with $\mathcal{S}(\bR^d)$ with $d\geq 1$. We feel that, in
the BMS setting such nomenclature may be confusing and we will not make use of
it.} to as $\left(L^2\right)_{\tilde\mch}$. 
}\\

\noindent Eventually we define \\
\definizione{\label{covdef}
A $\tgbms$ {\bf covariant field} is a section of the bundle\footnote{As in
the Poincar\'e invariant scenario, we implicitly assume that the following
continuous global section for the bundle (\ref{covbundle2}) has been chosen namely
$s:\mce^\prime\to\widetilde{G}$ mapping $\beta\mapsto\left(I,\beta\right)$ being
$I$ the identity element in $SL(2,\bC)$.}
$\mbH^\prime$ i.e. $\psi\in \left(L^2\right)_{\tilde\mch}$
which transforms under a unitary representation of the $\tgbms$ group as:
\beq\label{covariant}
\left[U(g)\psi\right](\beta)=e^{i(\beta,\alpha)}D(\Lambda)\psi(\Lambda^{-1}\beta),
\quad\forall g=(\Lambda,\alpha(\z,\bz))\in\tgbms
\eeq
where $D(\Lambda)$ is a unitary $SL(2,\bC)$ representation.
}\\

\noindent As in the induced scenario we shall work with a specific example namely:\\
\definizione{\label{scalarcov}
A $\tgbms$ {\bf real scalar covariant field} is a map $\psi$ which lies in
$\left(L^2\right)_{\bR}\equiv\left(L^2\right)$ which transforms as:
\beq\label{covbms}
\left[U(g)\psi\right](\beta)=e^{i(\beta,\alpha)}\psi(\Lambda^{-1}\beta).\quad\forall
g=\left(\Lambda,\alpha(\z,\bz)\right)\in\tgbms
\eeq
}

The definition \ref{covdef} (and consequently \ref{scalarcov}) is at this stage useless until two important aspects
are clarified. The first concerns the relation of (\ref{covariant}) with the 
induced wave function (\ref{induced}) which properly characterize a $\tgbms$ free field.
Following the seminal work of Wigner for the Poincar\'e group, such a problem has
been dealt with in \cite{Arcioni,DMP} where it has been shown that both
approaches, the induced and the covariant, are equivalent provided that suitable
constraints are imposed to (\ref{covariant}) in order to reduce it to
(\ref{induced}). Nonetheless such constraints should be interpreted as suitable
operators acting on $\left(L^2\right)_{\tilde\mch}$ and their definition requires the
introduction of a suitable space of test functions and of generalized functions
associated with $\left(L^2\right)_{\tilde\mch}$. The general theory has been developed
in the last twenty years and we refer to \cite{Kuo,Hida} for a detailed
discussion and for the proofs of the main statements. Conversely we will develop
now the construction for the specific scenario we are interested in. 

As a starting point and choosing for simplicity $\tilde\mch=\bC$ (or $\bR$ by a straightforward
adaption of the forthcoming analysis) , we recall that,
according to the It$\hat {\textrm o}$-Wiener theorem, each function $\psi\in\left(L^2\right)$ 
can be decomposed as 
\beq\label{Ito}
\psi(\beta)=\sum\limits_{n=0}^\infty I_n(f_n),\quad f_n\in C^\infty(\bS^2)_c^{\hat{
\otimes} n}
\eeq
where $C^\infty(\bS^2)_c^{\hat{\otimes}n}$ represents the complexification of the $n$-times 
symmetric tensor product of $C^\infty(\bS^2)$ whereas $I_n$ represents the multiple Wiener
integral defined as the linear functional $I_n:C^\infty(\bS^2)_c^{\hat{\otimes}n}
\to\bC$ such that for any $n_1+n_2+...=n$
\beq
I_n\left(\alpha_1(\z,\bz)^{\otimes n_1}\hat{\otimes}\alpha_2(\z,\bz)^{\otimes n_2}
\hat{\otimes}...\right)(\cdot)\negmedspace=\negmedspace\mcf_{n_1}\negmedspace\left[(\cdot,\alpha_1(\z,\bz))\right]
\negmedspace\mcf_{n_2}\negmedspace\left[(\cdot,\alpha_2(\z,\bz))\right]...,
\eeq 
where $\mcf_{n}\left[x\right]=(-)^n e^{\frac{x^2}{2}}\partial^n_x
e^{-\frac{x^2}{2}}$. 

A further interesting presentation of an element in $\left(L^2\right)$ consists of
showing that the It$\hat{\textrm o}$-Wiener decomposition (\ref{Ito}) is ultimately 
equivalent to the following sum (see chapter 5 in \cite{Kuo}):
\begin{equation}\label{Wick}
\psi(\beta)=\sum\limits_{n=0}^\infty\left(:\beta^{\otimes n}:, f_n\right)
\end{equation}
where $(,)$ refers to the canonical pairing between $C^\infty(\bS^2)$ and
$\mce^\prime$ whereas $:\beta^{\otimes n}:$ stands for the \emph{Wick tensor} 
$$:\beta^{\otimes
n}:=\sum\limits_{k=0}^{[n/2]}\binom{n}{2k}(2k-1)!!\beta^{\otimes(n-2k)}\widehat\otimes
\tau^{\otimes k},$$
where $\widehat\otimes$ is the symmetrized tensor product and
$\tau:\mce_c^{\otimes 2}\to\bC,$ is the trace operator mapping two elements $\eta,\xi$
in the complexification of $C^\infty(\bS^2)$ into
$$\left(\tau,\eta\otimes\xi\right)=\langle\eta,\xi\rangle,$$
being $\langle,\rangle$ the internal product in $L^2(\bS^2)$.

We may now state the following proposition \\
\proposizione{\label{Hidadistr}
Given the densely defined operator on $\left(L^2\right)$ $\Gamma(A)$ such that
$$\Gamma(A)\psi=\sum\limits_{n=0}^\infty I_n(A^{\otimes n}f_n),$$
then let us introduce for any $p\in\bN$ the set
\beq\label{EP}
\left(\mce\right)_p=\left\{\psi\in\left(L^2\right)\;\left|\right.\;\Gamma(A)^p\psi\in
\left(L^2\right)\right\}
\eeq
Closing $\left(\mce\right)_p$ to Hilbert space with respect to the norm 
$\left|\left|\psi\right|\right|_p=\left|\left|\Gamma(A)^p\psi\right|\right|_{\left(L^2\right)}$ then
we may introduce $\left(\mce\right)=\bigcap_p\left(\mce\right)_p$ as the 
projective limit of the sequence $\left(\mce\right)_p$ and 
$\left(\mce^\prime\right)_p$, $\left(\mce^\prime\right)$ respectively as the topological 
dual space of $\left(\mce\right)_p$ and of $\left(\mce\right)$. Then 
$\left(\mce\right)$ is a nuclear space with an associated Gelfand triplet 
$$\left(\mce\right)\subset\left(L^2\right)\subset\left(\mce^\prime\right),$$
and with the following series of continuous inclusions
$$\left(\mce\right)\hookrightarrow\left(\mce\right)_p\hookrightarrow\left(L^2\right)\hookrightarrow
\left(\mce^\prime\right)_p\hookrightarrow\left(\mce^\prime\right),$$
where $\left(\mce^\prime\right)_p$ is now the completion of
$\left(L^2\right)$ with respect to the norm $\left|\left|\psi\right|\right|_{-p}=
\left|\left|\Gamma(A)^{-p}\psi\right|\right|_{\left(L^2\right)}.$ The spaces
$\left(\mce\right)$ - endowed with the projective limit topology - and 
$\left(\mce^\prime\right)$ are respectively called the
{\bf space of Hida testing functionals and of Hida distributions}.}\\

The above proposition allows to identify a Gelfand triplet associated to the
space $\left(L^2\right)$ and thus we may refer to any element of
$\left(\mce\right)$ as a test function and of $\left(\mce^\prime\right)$ as a
distribution. We refer to $\langle\langle,\rangle\rangle$ as the natural pairing
between $\left(\mce\right)$ and $\left(\mce^\prime\right)$ subjected to the
compatibility condition that
\beq\label{compat2}
\langle\langle\psi(\beta),\psi^\prime(\beta)\rangle\rangle=\int\limits_{\mce^\prime}
d\nu(\beta)\psi(\beta)\psi^\prime(\beta),
\eeq
for any $\psi(\beta)\in\left(\mce\right)$ and for any
$\psi^\prime(\beta)\in\left(L^2\right)$.
Nonetheless, in order to correctly identify the constraints which
reduce the covariant to the induced wave function, we need now to introduce the
concepts of \emph{multiplication operator}. \\

\definizione{\label{mult2}
Given any $\alpha(\z,\bz)\in C^\infty(\bS^2)$, we call {\bf multiplication
operator} (along the $\alpha$-direction) the continuous operator 
$Q_\alpha:\left(\mce\right)\to\left(\mce\right)$ such that 
\beq\label{mult}
Q_\alpha\varphi(\beta)=(\beta,\alpha(\z,\bz))\varphi(\beta),\quad\forall\varphi\in\left(
\mce\right)\;\wedge\;\forall\alpha(\z,\bz)\in C^\infty(\bS^2)
\eeq
Furthermore we refer to $\widetilde Q_\alpha$ as the continuous extension of
$Q_\alpha$ to $\left(\mce^\prime\right)$ which is defined in analogy with 
(\ref{mult}).
}\\

Bearing in mind the above definitions we are now facing the following situation:
a $\tgbms$ covariant field (scalar or not) is, according to its definition and to
proposition \ref{Hidadistr}, a square integrable function over the space of
distributions over $\bS^2$ or, as well, a Hida testing functional if we take into
account that $\left(\mce\right)\subset\left(L^2\right)\subset\left(\mce^\prime\right)$. 

At the same time a $\tgbms$ free field is defined as in (\ref{sec1}) i.e. it is a square integrable
function whose support is a finite dimensional homogeneous space
$\mathcal{O}_\chi$ embedded in $\mce^\prime$.

We underline again that, according to Wigner seminal work for the Poincar\'e scenario, the above two
points of view are equivalent provided that suitable constraints are imposed on
the covariant field in order to reduce it to the induced counterpart.
In the BMS setting the overall idea is the same though we face a substantial
difference since, as we have outlined above, the covariant field has support on a
functional space and thus it is apparently rather counterintuitive that,
starting from a field $\psi\in L^2(\mce^\prime,\nu)\otimes\mch$ we shall find a
constraint reducing it to a function $\Psi\in
L^2(\mathcal{O}_\chi,\nu)\otimes\mch^\prime$. We have already addressed this
problem in \cite{DMP} though not in the rigorous frame of Hida distributions. We
will now provide a constructive demonstration of Wigner idea for the specific
scenario of a real scalar field with mass $m$ i.e. $\mch=\mch^\prime=\bR$ and 
the orbit of the induced wave function is the hyperboloid $\frac{SL(2,\bC)}{SU(2)}$ 
if $m^2> 0$ or $\frac{SL(2,\bC)}{\Delta}$ if $m^2=0$.

The starting point consists of introducing a finite dimensional counterpart of
the elements in $\left(\mce\right)^\prime$. We will state now some results first
appeared in \cite{Kubo} and here stated in our specific scenario. Adaption to
the general scenario is straightforward.\\

\definizione{\label{finite}
Let $C^\infty(\bS^2)\subset\mch\subset\mce^\prime$ be the Gelfand triplet 
constructed in remark \ref{triplet} out of which the space
of Hida distributions $\left(\mce^\prime\right)$ has been constructed as in
proposition \ref{Hidadistr}. Then if we choose any $k$-tuple
$\left\{e_1,...e_k\right\}\subset L^2(\bS^2)$ with $k<\infty$ and if we refer to $V$ 
as the real linear space spanned by $e_1,...,e_k$, we may introduce the space
$\left(\mce^\prime\right)_V$ as the $\left(\mce^\prime\right)$-closure of all
polynomials in $\langle\cdot,\vec
e\rangle\doteq\left(\langle\cdot,e_1\rangle,...,\langle\cdot,e_k\rangle\right)$.
Then we call $\psi\in\left(\mce^\prime\right)$ a {\bf finite dimensional Hida
distribution} if $\psi\in\left(\mce^\prime\right)_V$ for some finite dimensional
subspace $V$ constructed as above. We call $\left(\mce\right)_V\doteq\left(\mce\right)
\cap\left(\mce^\prime\right)_V$ the space of {\bf finite dimensional Hida test
functions}.
}\\

The above definition clearly underlines that certain specific Hida distributions/testing functionals 
could be interpreted as finite dimensional distributions/testing functionals. The natural subsequent
step would be to interpret them as Schwartzian generalized functions or testing functionals over $\bR^k$ though it
is rather straightforward to realize that the Gelfand triplet
$\mathcal{S}(\bR^k)\subset L^2(\bR^k)\subset\mathcal{S}^\prime(\bR^k)$ does not
fit in this picture since a priori there is no reason why a finite dimensional
Hida distribution should lie in the dual space of rapidly decreasing test
functions. Thus we need to introduce a new auxiliary Gelfand triplet; the
starting point consists in $\mathcal{P}(\bR^k)$ which is the space of
polynomials in $x_\mu=(x_1,...,x_k)\in\bR^k$ with $k<\infty$. Referring to
$\mu_k$ as the standard Gaussian measure on $\bR^k$, we may close
$\mathcal{P}(\bR^k)$ to Hilbert space - say $\overline{\mathcal{P}(\bR^k)}$ -  
with respect to the inner product
$$\left(F,G\right)=\int\limits_{\bR^k}F(x_\mu)G(x_\mu)d\mu_k(x_\mu).\quad\forall
F,G\in \mathcal{P}(\bR^k)$$ 
We shall construct a Gelfand triplet out of this Hilbert space considering the
Ornstein-Uhlenbeck operator on $\bR^k$ i.e. $L=\nabla-\sum\limits_{i=1}^k
x_i\frac{\partial}{\partial x_i}$, where $x_i$ are the Cartesian coordinates on
$\bR^k$ whereas $\nabla$ is the Laplacian operator on
$\bR^k$. As shown in \cite{Kubo} we can now exploit proposition (\ref{prima})
with respect to the basis for $L$ in $\overline{\mathcal{P}(\bR^k)}$ given by 
the vector $H_{\bf n}(x_\mu)=\prod\limits_{i=1}^k H_{\bf n_i}(x_i)$ being 
$H_{\bf n_i}$ the $n_i$-th Hermite polynomial. Thus we introduce the sequence of 
spaces - with respect to the parameter $t\in\bR$ -
\begin{equation}
\mathcal{I}_t(\bR^k)=\left\{\psi\in\overline{\mathcal{P}(\bR^k)}\;\left|\right.\;
\exp{(-tL)}\psi\in\overline{\mathcal{P}(\bR^k)}\right\}.
\end{equation}
Closing this space to Hilbert space with respect to the internal product
$$\left(F,G\right)_t=\left(\exp{(-tL)}F,\exp{(-tL)}G\right),\quad\forall F,G\in
\mathcal{I}_t(\bR^k),$$
one ends up for any real positive value of $t$ with the sequence of continuous
inclusions
$$\mathcal{I}_t(\bR^k)\hookrightarrow\overline{\mathcal{P}(\bR^k)}
\hookrightarrow\mathcal{I}_{-t}(\bR^k).$$
Considering now the projective limit space
$\mathcal{I}(\bR^k)=\bigcap\limits_t\mathcal{I}_t(\bR^k)$, endowed as in
proposition \ref{prima} with the projective limit topology, we may construct the
new Gelfand triplet
\begin{equation}\label{newGelfand}
\mathcal{I}(\bR^k)\subset\overline{\mathcal{P}(\bR^k)}\subset\mathcal{I}^\prime
(\bR^k),
\end{equation}
where $\mathcal{I}^\prime(\bR^k)$ is the topological dual space of $\mathcal{I}
(\bR^k)$.

We can now formulate a characterization theorem for finite dimensional Hida
testing functionals whose demonstration has been given in \cite{Kubo} and which is here
stated in terms of our specific framework:\\

\teorema{\label{finiteHida}
Referring to the Gelfand triplet  $C^\infty(\bS^2)\subset L^2(\bS^2)\subset \mce^\prime$
let us choose, as in definition \ref{finite}, a $k$-tuple 
$\left\{e_i\right\}_{i=1}^k\in C^\infty(\bS^2)$ whose elements are mutually orthogonal
with respect to the inner product in $L^2(\bS^2)$ and let us call $V=span\left\{e_i
\right\}_{i=1}^k$. Then for any finite dimensional Hida testing functional $\psi\in
\left(\mce\right)_V$, it exists a function $F\in\mathcal{I}(\bR^k)$ such
that $\varphi=F(x_\mu)$ where $x_\mu=\left(\langle\cdot,e_1\rangle,...,\langle
\cdot,e_k\rangle\right)$.

Furthermore let us introduce the projector $\pi_V:\mce^\prime\to V$ which maps
$y\in \mce^\prime$ to $\pi_V(y)=\sum\limits_{i=1}^k\left(y,e_i\right)e_i$ where $(,)$
stands for the pairing between $C^\infty(\bS^2)$ and $\mce^\prime$. Then
$\pi_V$ automatically induces a projection operator $\Pi_V:\left(\mce^\prime\right)\to
\left(\mce^\prime\right)_V$ such that $\Pi_V\doteq\Gamma(\pi_V)$ maps any 
$\psi\in\left(\mce^\prime\right)$ in
\begin{equation}\label{defining2}
\Pi_V\psi=\Pi_V\left(\sum\limits_{n=0}^\infty
I_n(f_n)\right)=\sum\limits_{n=0}^\infty I_n(\pi_v^{\otimes n}f_n),
\end{equation}
being $I_n$ the multiple Wiener integral as in (\ref{Ito}). 
Thus we conclude that $\psi\in\left(\mce^\prime\right)_V$ iff
\begin{equation}\label{Vproj}
\Pi_V\psi=\psi.
\end{equation}}

The above theorem grants us that any Hida testing functional which satisfies
(\ref{Vproj}) naturally identifies a function lying in $\mci(\bR^k)$; this is not 
the answer we were looking for since we ultimately seek an element at least in
$\mathcal{S}(\bR^k)\subset L^2(\bR^k)$. Thus we need to exploit another theorem proved in
\cite{Kubo} and here adapted to our specific scenario:\\

\proposizione{\label{Lesbegue}
If a function $F(x_\mu)$ lies in $\mci(\bR^k)$ then
$F(x_\mu)e^{-\frac{1}{4}\delta^{\mu\nu}x_\mu x_\nu}$ lies in
$\mathcal{S}(\bR^k)$ being $\delta^{\mu\nu}$ the Kroneker delta.}\\

We have now all the ingredient to exploit the theory of finite-dimensional Hida
distributions to construct the equations of motion for the BMS free field. 
The first step consists of remembering that both the orbit of the massive and
massless real scalar $\tgbms$ field lies in $\left(T^4\right)^\prime$. Bearing
in mind that such a space is generated by real linear combinations out of the
basis $\left\{Y^*_{lm}\right\}_{l=0}^1\subset\mce^\prime$ defined as
$\left(Y^*_{lm},Y_{l'm'}(\z,\bz)\right)=\delta_{ll'}\delta_{mm'}$, it is natural to
choose $V=T^4=\left\{Y_{lm}(\z,\bz)\right\}_{l=0,1}$.\\

\begin{lemma}\label{eq1}
The orbit/support of a covariant real (massive or massless) scalar field $\psi$ lies in $(T^4)^\prime$ 
iff  
\begin{equation}\label{projorbit}
\Pi_{T^4}\psi(\beta)=\psi(\beta).
\end{equation}  
\end{lemma}

\emph{Proof.}
We exploit the It$\hat{\textrm o}$-Wiener decomposition of a generic functional in
$\left(L^2\right)$ as $\psi(\beta)=\sum\limits_{n=0}^\infty\left(:\beta^{\otimes
n}:,f_n\right)$ and (\ref{defining2}). According to this latter equation and
introducing $e_\mu=\left(Y_{00}(\z,\bz),...,Y_{11}(\z,\bz)\right)$, (\ref{projorbit}) 
reads:
$$\sum\limits_{n=0}^\infty\left(:\beta^{\otimes
n}:,\sum\limits_{\mu=0}^4\left(e_\mu,f_n\right)e_\mu\right)=\sum\limits_{n=0}^\infty
\left(:\beta^{\otimes n}:,f_n\right),$$
which is satisfied iff $\beta\in\left(T^4\right)^\prime$ or $f_n$ lies in
$\left(T^4\right)^{\widehat\otimes n}$ for any $n$. In this latter case we shall make use
of proposition \ref{annihil} - more precisely of the considerations in its proof - to
conclude that, whenever a generic distribution $\beta\in\mce^\prime$ is evaluated with
a test function $\alpha(\z,\bz)\in T^4$, this is equal to extract from $\beta$ a 
representative in an equivalence class of $\frac{\mce^\prime}{\left(T^4\right)^0}$
and evaluate it with $\alpha(\z,\bz)$.
Bearing now in mind the $SL(2,\bC)$-invariant isomorphism between 
$\frac{\mce^\prime}{\left(T^4\right)^0}$ with $\left(T^4\right)^\prime$, the statement
of the theorem is naturally implied. $\Box$\\

For later convenience it is interesting to notice at this stage that the above 
equation of motion can be also written in terms of operators acting on the covariant 
wave function namely, referring to definition \ref{mult2}, the following
lemma holds:\\

\lemma{\label{eq3}
Bearing in mind the decomposition in theorem \ref{decompos}, a field 
$\psi\in\left(L^2\right)$ satisfies (\ref{projorbit}) iff 
\begin{equation}\label{constr}
Q_{\alpha(\z,\bz)}\psi(\beta)=0.\quad\forall\alpha(\z,\bz)\in ST
\end{equation}}

\emph{Proof.}
According to definition \ref{mult2},
$Q_\alpha(\z,\bz)\psi(\beta)=\left(\beta,\alpha(\z,\bz)\right)\psi(\beta)$; it
is immediate to realize that if (\ref{projorbit}) holds, then lemma \ref{eq1}
grants us that $\beta$ can be chosen in $\left(T^4\right)^\prime$ and, unless
$\psi(\beta)$ is identically vanishing, (\ref{constr}) is zero iff
$\left(T^4\right)^\prime\subseteq\left(ST\right)^0$ which is the annihilator of
$ST$. At the same time if we suppose that (\ref{constr}) holds then
$\beta\in\left(ST\right)^0$ and (\ref{projorbit}) holds iff $\left(ST\right)^0
\subseteq\left(T^4\right)^\prime$. We need only to demonstrate that it exists an
isomorphism between $\left(T^4\right)^\prime$ and $\left(ST\right)^0$.

The starting point consists of exploiting theorem \ref{decompos} according to
which the factor space $\frac{C^\infty(\bS^2)}{ST}$ is isomorphic to the
subspace $T^4\subset C^\infty(\bS^2)$. Accordingly, per duality, also
$\left(T^4\right)^\prime$ is isomorphic to
$\left(\frac{C^\infty(\bS^2)}{ST}\right)^\prime$. Furthermore any
$\beta\in\left(T^4\right)^\prime$ can be extended according to theorem \ref{annihil}
to a functional $\tilde\beta$ on $\mce^\prime$ in such a way that, given any two
$\alpha(\z,\bz),\alpha^\prime(\z,\bz)\in C^\infty(\bS^2)$,
$\tilde\beta\left(\alpha(\z,\bz)\right)=\tilde\beta\left(\alpha^\prime(\z,\bz)
\right)$ if $\alpha(\z,\bz)-\alpha^\prime(\z,\bz)\in ST$. Per linearity of the
elements in $\mce^\prime$, it implies $\tilde\beta\left(\alpha(\z,\bz)-\alpha^\prime
(\z,\bz)\right)$ vanishes i.e. $\tilde\beta\in \left(ST\right)^0$ and
$\left(T^4\right)^\prime\subseteq\left(ST\right)^0$. 

To show the opposite inclusion let us start from any
$\beta\in\left(ST\right)^0$. We can now exploit theorem \ref{decompos} according
to which $ST$ is a subspace of $C^\infty(\bS^2)$ and thus, according to theorem
\ref{annihil}, $\beta$ can be extended to a functional in $\mce^\prime$. Choose
any such extension - say $\tilde\beta$ - and evaluate it on any $\alpha(\z,\bz)\in
C^\infty(\bS^2)$. Still according to theorem \ref{decompos}, $\alpha(\z,\bz)$
can be univocally split in the sum of $\alpha^\prime(\z,\bz)\in T^4$ and
$\tilde\alpha(\z,\bz)\in ST$. Thus, per linearity,
$$\tilde\beta(\alpha(\z,\bz))=\tilde\beta(\alpha^\prime(\z,\bz))+\tilde\beta(
\tilde\alpha(\z,\bz))=\tilde\beta(\alpha^\prime(\z,\bz)),$$
where the last equality holds since $\tilde\beta$ must agree with $\beta$ on
$ST$. Thus the above equation grants us that
$\tilde\beta\in\left(T^4\right)^\prime$ i.e.
$\left(ST\right)^0\subseteq\left(T^4\right)^\prime$, which concludes the demonstration.
$\Box$\\

We have now identified the class of covariant $\tgbms$ scalar fields $\psi$ which are 
supported on $\left(T^4\right)^\prime$. The last step consists of choosing 
suitable constraints which grant us that $\psi$ is supported either on the
hyperboloid $\frac{SL(2,\bC)}{SU(2)}$ either on the light cone 
$\frac{SL(2,\bC)}{\Delta}$. From a physical perspective this amounts to assign a fixed
value for the mass to the covariant field and, from an operative point of view,
it translates in the following lemma:\\

\lemma{\label{eq2}
A $\tgbms$ covariant scalar field $\psi$ has support on the orbit generated by
$\frac{SL(2,\bC)}{SU(2)}$ action on $\bar\beta_1=\sqrt{\frac{3}{4\pi}}mY^*_{00}$ or on
that generated by $\frac{SL(2,\bC)}{\Delta}$ action on $\bar\beta_2=C\delta$ iff,
besides (\ref{projorbit}), $\psi$ satisfies 
\begin{equation}\label{KG}
\eta^{\mu\nu}Q_{e_\mu}Q_{e_\nu}\psi(\beta)=\left\{\begin{array}{ll}
0 &\quad\textrm{for the little group}\; \Delta\\
m^2\psi(\beta)&\quad\textrm{for the little group}\; SU(2)
\end{array}\right.,
\end{equation}
where $e_\mu=\left(Y_{00}(\z,\bz),...,Y_{11}(\z,\bz)\right)$ is the 4-vector of elements in 
$C^\infty(\bS^2)$ and $\eta^{\mu\nu}=diag(-1,1,1,1)$. }\\

\emph{Proof.}
Suppose that $\psi(\beta)$ is supported on the orbit generated by
$\frac{SL(2,\bC)}{L_i}$ on $\bar\beta_i$ where $i=1,2$ and $L_1=SU(2)$ and
$L_2=\Delta$. Then, for any point $\beta$ on one of the two orbit, it exists $\Lambda\in
SL(2,\bC)$ such that $\beta=\Lambda\bar\beta_i$ and the following chain of identities
holds:
$$\eta^{\mu\nu}Q_{e_\mu}Q_{e_\nu}\psi(\beta)=\eta^{\mu\nu}Q_{e_\mu}Q_{e_\nu}\psi(\Lambda
\bar\beta_i)=\eta^{\mu\nu}\left(e_\mu,\Lambda\bar\beta_i\right)\left(e_\nu,\Lambda\bar
\beta_i\right)\psi(\Lambda\bar\beta_i)=$$
$$=B(\Lambda\bar\beta_i,\Lambda\bar\beta_i)
\psi(\Lambda\bar\beta_i)=m^2\psi(\beta),$$
where $m^2$ is either $0$ or different from $0$ depending on the chosen little group.
In the above chain of identities we have exploited the multiplication
operator as introduced in definition \ref{mult2} whereas, in the last two identities,
we refer to proposition \ref{mass2} and, in particular, to the definition of the real bilinear form
(\ref{mass}) and its $SL(2,\bC)$ invariance.

The converse is rather straightforward. Suppose a covariant scalar $\tgbms$ field
satisfies (\ref{projorbit}). Then $\beta\in\left(T^4\right)^\prime$ and (\ref{KG}) becomes
$$\left[\eta^{\mu\nu}(\beta,e_\mu)(\beta,e_\nu)-m^2\right]\psi(\beta)=0.\quad\beta\in
\left(T^4\right)^\prime$$
Thus, unless $\psi(\beta)$ is identically vanishing, $\eta^{\mu\nu}(\beta,e_\mu)(
\beta,e_\nu)-m^2=0$ which is, depending on the chosen value for $m^2$,
the defining equation for the mass hyperboloid or for the light cone realized in $\bR^4$. 
We need at last to show
that the orbit is necessarily generated by the fixed point $\bar\beta_i$. This is still
straightforward; suppose that $m^2\neq 0$, then we just need to exploit that $\eta^{
\mu\nu}(\beta,e_\mu)(\beta,e_\nu)$ is the $SL(2,\bC)$ invariant bilinear form
$B(\beta,\beta)$ as in (\ref{mass}). Thus we may find $\Lambda\in SL(2,\bC)$ such that
$B(\beta,\beta)=B(\Lambda\beta,\Lambda\beta)=m^2$ and
$B(\Lambda\beta,\Lambda\beta)=(\beta^\prime,e_0)(\beta^\prime,e_0)$ where
$\beta^\prime=\Lambda\beta$. Since $e_0=Y_{00}(\z,\bz)$, $\beta^\prime$ should be
equal to a constant times $Y^*_{00}$ plus a term lying in the annihilator\footnote{The
reader should bear in mind that the annihilator of any $e_\mu\in T^4$ is the set of
elements $f$ in $\left(T^4\right)^\prime$ such that $f(ke_\mu)=0$ for any $k\in\bR$.} 
of $e_0$ - say $(Y_{00})^0$. To be rigorous one now should exploit proposition 
\ref{annihil} to show that it exists an isomorphism between 
$\frac{(T^4)^\prime}{(Y_{00})^0}$ and the space dual to one dimensional subspace of 
$T^4$ generated by $Y_{00}$. Thus one can always choose the representative in such 
factor group in such a way that it coincides with $\bar\beta_1$ i.e. the distribution 
generating the orbit for the massive canonical $\tgbms$ scalar field. An identical 
procedure leads to the same conclusion for the massless case and thus the statement is 
proved. $\Box$ \\

We have almost completed our task. According to lemma \ref{eq1} and \ref{eq2} we have
shown that a $\tgbms$ covariant scalar field $\psi\in\left(L^2\right)$ satisfying
(\ref{covbms}) can be reduced to a function on the mass hyperboloid or on the light
cone transforming under a scalar $\tgbms$ unitary and irreducible representation
(respectively induced from the $SU(2)$ and the $\Delta$ subgroups of $SL(2,\bC)$) iff
it satisfies the equations (\ref{projorbit}) and (\ref{KG}). 

The tricky point is the following: can we conclude that this function is square
integrable with respect to the measure on each orbit? At this stage this is definitely not
possible since theorem \ref{finiteHida} grants us that a map $\psi\in\left(L^2\right)$
such that $\Pi_{T^4}\psi(\beta)=\psi(\beta)$ is in one to one correspondence with the
functions in $\mci(\bR^4)$ - say $\psi(p_\mu)$ with $p_\mu=<\beta,e_\mu>$ - which is 
is continuously embedded in the Hilbert space $\overline{\mathcal{P}(\bR^4)}$ which we 
remember being the Hilbert space of polynomial function with respect to the canonical 
Gaussian measure on $\bR^4$. Thus, from one side this inclusion justifies the claim 
that the covariant field satisfying (\ref{projorbit}) and (\ref{KG}) transforms under 
a unitary $\tgbms$ induced representation whereas from the other side it allow us to 
exploit theorem \ref{Lesbegue} to claim that
$\tilde\psi(p_\mu)=e^{-\delta^{\mu\nu}p_{\mu}p_{\nu}}\psi(p_\mu)$ lies in 
$\mathcal{S}(\bR^4)$ i.e. $\tilde\psi(p_\mu)$ is square-integrable with respect to the
Lesbegue measure on $\bR^4$. The remaining constraint (\ref{KG}) does not harm the
previous reasoning since it corresponds in the space $\mci(\bR^4)$ to impose the
usual equation 
$$\left[\eta^{\mu\nu}p_\mu p_\nu-m^2\right]\psi(p_\mu)=0,$$
which is also identically satisfied by $\tilde\psi(p_\mu)=0$. Thus we can summarize
the full construction in the following theorem:\\

\teorema{\label{inizioditutto}
A covariant $\tgbms$ (massive or massless) scalar field $\psi:\mce^\prime\to\bR$ which
transforms as (\ref{covbms}) and which satisfies the equations (\ref{projorbit}) and 
(\ref{KG}) corresponds to a $\tgbms$ induced scalar field ((\ref{massive}) or
(\ref{massless})) up to the rescaling of the latter by $e^{-\delta^{\mu\nu}p_\mu
p_\nu}$.}\\

\remark{
The construction outlined above refers to the special case of the scalar fields. In
particular, for the massless case, we referred to an induced wave function living on a 
rather specific orbit. The real purpose behind such a choice arises from a physical
perspective since, as we have outlined before, up to now the $\tgbms$ fields on null infinity
which can be physically 
interpreted from an holographic point of view are those supported on
$\left(T^4\right)^\prime$ (see for example \cite{Dappiaggi,DMP}). Nonetheless the
overall idea for the above construction can be slavishly applied to a generic $\tgbms$
covariant field in order to reduce it to its induced counterpart. The real tricky
issue would be to construct case by case the suitable constraints and in particular to
select a specific set of orthonormal functions in $L^2(\bS^2)$ out of which construct
a finite dimensional Hida testing functional starting from the whole covariant field.
}

\subsection{$\tgbms$ equations of motion as evolution equations}
To conclude this section it is natural to deal with the following remark:
the equations of motion for a $\tgbms$ (massless or massive) scalar field 
are constraint equations in direct analogy with the counterpart in a Poincar\'e
invariant free field in the momenta space.
Nonetheless it is often more convenient to deal either in classical either in quantum
field theory with an evolution problem i.e. (at least) a partial differential
equation. In order to switch to this perspective, in our scenario, we need to
introduce two key ingredients: a differential operator on the space of Hida testing
functionals and distributions and a suitable notion of  ``Fourier-like'' transform 
$\mathcal{F}$. The answer to this query has been developed and extensively
discussed in \cite{Kuo,Hida} and we will limit ourselves to the main
definitions:\\

\definizione{\label{deriv}
Let us consider any Hida testing functional $\psi(\beta)$ on $\left(\mce\right)$; 
we define the Gateaux derivative of $\psi(\beta)$ along the direction
$\tilde\beta\in\mce^\prime$ as the continuous operator $\mcd_{\tilde\beta}:
\left(\mce\right)\to\left(\mce\right)$ such that
\beq\label{Gateaux}
\mcd_{\tilde\beta}\psi(\beta)=\lim\limits_{\epsilon\to
0}\frac{\psi(\beta+\epsilon\tilde\beta)-\psi(\beta)}{\epsilon}=\sum\limits_{n=1}
^\infty\left(:\beta^{\otimes (n-1)}:,(\tilde\beta,f_n)\right)
\eeq
The operator $\mcd_{\tilde\beta}$ admits a unique continuous extension to an
operator $\widetilde\mcd_{\tilde\beta}:\left(\mce^\prime\right)\to\left(\mce^\prime
\right)$.
}\\

We are going to state now an important result which relates the multiplication
operator with the Gateaux derivative. The following lemma, proved in \cite{Kuo},
also shows that, opposite to the usual behaviour such as on the space of 
Schwartz test functions, the multiplication operators is a sort of ``derivative 
operator'' i.e. it obeys a Liebnitz rule.\\

\lemma{\label{DQ}
For any $\alpha(\z,\bz)\in C^\infty(\bS^2)$, the following equality holds:
$$Q_{\alpha(\z,\bz)}=\mcd_{\alpha(\z,\bz)}+\mcd^*_{\alpha(\z,\bz)},$$
which is meant as a continuous operator form $\left(\mce\right)$ into itself.
Furthermore, for any $\beta\in\mce^\prime$, it also holds:
$$Q_\beta=\mcd_\beta+\mcd^*_\beta,$$
which is meant as a continuous operator from $\left(\mce\right)$ into
$\left(\mce^\prime\right)$.}\\

In order to switch from a constraint equation such as (\ref{constr}) and (\ref{KG}) to
an evolution equation, the natural step in the canonical formulation of quantum field
theory over Minkowski background consists of performing a Fourier transform $\tilde\mcf$.
In this latter framework such a transformation is a continuous linear operator from
the space of Schwartz test function $\mathcal{S}(\bR^d)$ ($d\geq 1$) into itself which
can also be defined on the dual space $\mathcal{S}^\prime(\bR^d)$ as the adjoint
operator $\tilde\mcf^*$. 

On the opposite, in the framework of white noise analysis, the definition of 
Fourier transform is instead a little less intuitive and its main peculiarity lies in
the fact that it is constructed only as an operator from the space of Hida 
distributions - $\left(\mce^\prime\right)$ - into itself. 
Nonetheless, since it
represents a key component for the analysis of $\tgbms$ covariant free fields
and of their equations of motion we shall now introduce it following chapter 11 of
\cite{Kuo}\\
\definizione{
We call {\bf S-transform} of $\Psi\in\left(\mce^\prime\right)$ the functional
$\Psi^S:C^\infty(\bS^2)\to\bC$ such that
$$\Psi^S\left(\alpha(\z,\bz)\right)=\langle\langle\Psi,:e^{\left(\cdot,\alpha(\z
,\bz)\right)}:\rangle\rangle,$$
where $\alpha(\z,\bz)\in C^\infty(\bS^2)_{\bC}$ and
$\langle\langle,\rangle\rangle$ stands for the pairing between
$\left(\mce\right)$ and $\left(\mce^\prime\right)$. 

We call {\bf Fourier transform} the continuous linear operator 
$\mcf:\left(\mce^\prime\right)\to\left(\mce^\prime\right)$ such that
$\widehat\Psi\doteq\mcf\Psi$ is the unique element in $\left(\mce^\prime\right)$
satisfying
\begin{equation}
\widehat\Psi^s\left(\alpha(\z,\bz)\right)=\langle\langle\Psi,e^{-i\left(\cdot,
\alpha(\z,\bz)\right)}\rangle\rangle=\Psi^s(-i\alpha(\z,\bz))e^{-\frac{1}{2}
\left|\left|\alpha(\z,\bz)\right|\right|_{L^2}},
\end{equation}
where $\left|\left|,\right|\right|_{L^2}$ stands for the norm in $L^2(\bS^2)$.

Equivalently the Fourier transform is the unique continuous linear operator from
$\left(\mce^\prime\right)$ into itself such that for any $\alpha(\z,\bz)\in
C^\infty(\bS^2)$.
\beq\label{multdiff}
\mcf\widetilde\mcd_{\alpha(\z,\bz)}=i\widetilde Q_{\alpha(\z,\bz)}\mcf,\quad
\mcf\widetilde Q_{\alpha(\z,\bz)}=i\widetilde\mcd_{\alpha(\z,\bz)}\mcf,
\eeq
where $\widetilde\mcd$ is defined as in definition \ref{deriv} and $\widetilde Q$
as in definition \ref{mult2}.
}

The above definition apparently put us into the position to rewrite in terms of 
derivative operators the equations of motion for a $\tgbms$ massive or massless 
scalar field which, according to lemma \ref{eq1}, \ref{eq3} and \ref{eq2}, are 
given by  (\ref{constr}) and (\ref{KG}). Interpreting the field 
$\psi\in\left(\mce\right)$ as an element in $\left(\mce^\prime\right)$ by means of the
continuous inclusion of $\left(\mce\right)$ in $\left(\mce^\prime\right)$, these latter
equations become:
\begin{gather}\label{fake1}
\widetilde\mcd_{\alpha(\z,\bz)}\widehat{\psi}=0,\quad\forall\alpha(\z,\bz)\in ST\\
\eta^{\mu\nu}\widetilde\mcd_{e_\mu}\widetilde\mcd_{e_\nu}\widehat\psi=\left\{
\begin{array}{ll}
0 &\quad\textrm{for the little group}\; \Delta\\
m^2\widehat\psi&\quad\textrm{for the little group}\; SU(2)
\end{array}\right.,\label{fake2}
\end{gather}
where $\widehat\psi\in\left(\mce^\prime\right)$. 

It is imperative to underline a key aspects of the above differential equations:
although (\ref{fake2}) is similar to the well-known Klein-Gordon equation in Minkowski
spacetime, $\eta^{\mu\nu}\widetilde\mcd_{e_\mu}\widetilde\mcd_{e_\nu}$, is not a 
symmetric operator on $\left(\mce^\prime\right)$ contrary to
$\eta^{\mu\nu}\partial_\mu\partial_\nu$ on $\mcs(\bR^d)$. This difference is rather
important since, as we shall see in the next section, if we wish to interpret the
equations of motion of our fields as the extremum of a suitable (Lagrangian)
functional, than the defining operator must be symmetric.

To avoid such a problem, we shall now exploit a new kind of
transformation which has been discussed in \cite{Kuo,Lee}. Still referring to our
specific scenario the following holds: \\

\definizione{\label{tran}
We call {\bf Fourier-Gauss} transform the continuous linear operator
$\mcg_{a,b}:\left(\mce\right)\to\left(\mce\right)$ such that, being
$a,b\in\bC-\left\{0\right\}$,
\beq\label{FG}
\mcg_{a,b}\psi(\beta)=\int\limits_{\mce^\prime}\psi(a\beta^\prime+b\beta)d\mu(
\beta^\prime).\quad\forall\psi\in\left(\mce\right)
\eeq
Furthermore, for all $a,b\in\bC-\left\{0\right\}$ and for all $\alpha(\z,\bz)\in C^\infty(\bS^2)$, it 
holds that 
\begin{gather}
\mcg_{a,b}\mcd_{\alpha(\z,\bz)}=b^{-1}\mcd_{\alpha(\z,\bz)}\mcg_{a,b},\label{uno}\\
\mcg_{a,b}Q_{\alpha(\z,\bz)}=a^2b^{-1}\mcd_{\alpha(\z,\bz)}\mcg_{a,b}+bQ_{\alpha(\z,
\bz)}\mcg_{a,b},\label{due}
\end{gather}
where $\mcd_\eta$ and $Q_\eta$ are the derivative and multiplication operators
respectively introduced in definition \ref{deriv} and \ref{mult2}. Moreover, bearing
in mind the spaces $\left(\left(\mce\right)_p,\left|\left|,\right|\right|_p\right)$ as
introduced in (\ref{EP}), if
$a^2+b^2=1$ and $|b|=1$, then $\left|\left|\mcg_{a,b}\psi(\beta)\right|\right|_p=
\left|\left|\psi(\beta)\right|\right|_p$ for all $\psi\in\left(\mce\right)_p$ and for
all $p\geq 0$.
}\\

We seek now to single out a preferred $\mcg_{a,b}$ within the set of Fourier-Gauss 
transforms parametrized by the complex numbers $a,b$. The criterion, we shall refer
to, consists of requiring that the kernel of the operator 
$\eta^{\mu\nu}\mcd_{e_\mu}\mcd_{e_\nu}$ is mapped into the kernel of a new but 
symmetric operator. Bearing in mind that, for a linear operator, such a condition
coincides with the request of self-adjointness, the following proposition holds:\\

\proposizione{\label{propFG}
There are only two Fourier-Gauss transforms, namely $\mcg_{\sqrt{2},i}$ and
$\mcg_{\sqrt{2},-i}$, such that $\mcg_{a,b}\eta^{\mu\nu}Q_{e_\mu}Q_{e_\nu}=A(a,b)
\mcg_{a,b}$ where $A(a,b)$ is a linear continuous selfadjoint operator on $\left(\mce
\right)$ which admits an extension to a unitary operator on $\left(L^2\right)$. 
Furthermore 
$$A(\sqrt{2},\pm i)=\pm i\eta^{\mu\nu}\left(Q_{e_\mu}-2\mcd_{e_\mu}
\right)\left(Q_{e_\nu}-2\mcd_{e_\nu}\right),$$
where the plus stands for $b=-i$ whereas the minus for $b=i$. }\\

\emph{Proof.}
According to definition \ref{tran}, for any non vanishing $a,b\in\bC$ the
Fourier-Gauss transform is a continuous linear operator from $\left(\mce\right)$ into
itself such that, exploiting (\ref{uno}), 
$$\mcg_{a,b}\eta^{\mu\nu}Q_{e_\mu}Q_{e_\nu}=\eta^{\mu\nu}\left[a^4b^{-2}
\mcd_{e_\mu}\mcd_{e_\nu}+\right.$$
$$+b^{-2}Q_{e_\mu}Q_{e_\nu}+\left.a^2\left(\mcd_{e_\mu}Q_{e_\nu}+Q_{
e_\mu}\mcd_{e_\nu}\right)\right]\mcg_{a,b}.$$
In order to realize when 
$$A(a,b)=\eta^{\mu\nu}\left[a^4b^{-2}\mcd_{e_\mu}\mcd_{e_\nu}+
b^{-2}Q_{e_\mu}Q_{e_\nu}+a^2\left(\mcd_{e_\mu}Q_{e_\nu}+Q_{e_\mu}\mcd_{e_\nu}\right)
\right]$$
is self-adjoint on $\left(\mce\right)$ we refer to lemma \ref{DQ} and to the
relation $Q^*_{\alpha(\z,\bz)}=Q_{\alpha(\z,\bz)}$ on $\left(\mce\right)$ for any
$\alpha(\z,\bz)\in C^\infty(\bS^2)$ according to which:
$$A^*(a,b)=\eta^{\mu\nu}\left[a^4b^{-2}\mcd_{e_\mu}\mcd_{e_\nu}+\left(a^4b^{-2}+b^2+
2a^2\right)Q_{e_\mu}Q_{e_\nu}+\right.$$
$$\left.-\left(a^4b^{-2}+a^2\right)\left(Q_{e_\mu}\mcd_{e_\nu}+
\mcd_{e_\mu}Q_{e_\nu}\right)\right].$$
Thus, on $\left(\mce\right)$, $A^*(a,b)=A(a,b)$ iff $a^2=-2b^2$. If we require that 
$\mcg_{a,b}$ could also be extended to a unitary operator on $\left(L^2\right)$, then,
according to definition \ref{tran}, we also impose $\left|b\right|=1$ and $a^2+b^2=1$; it
implies that $b=\pm i$ and $a=\pm\sqrt{2}$.

To conclude we refer to theorem 11.28 in \cite{Kuo} and remarks below according to
which $\mcg_{a,b}=\mcg_{c,d}$ iff $a=\pm c$ and $b=d$. Thus $\mcg_{\sqrt{2},\pm
i}=\mcg_{-\sqrt{2},\pm i}$ on $\left(\mce\right)$. $\Box$\\

\remark{
The arbitrariness in the choice of the Fourier-Gauss transform which arises from the
previous theorem is only apparent. If we exploit theorem 11.30 in \cite{Kuo},
according to which, for any $a,b,c,d\in\bC-\left\{0\right\}$,
$\mcg_{c,d}\mcg_{a,b}=\mcg_{\pm\sqrt{a^2+b^2c^2},bd}$, we end up with
$$\mcg_{\sqrt{2},i}=\mcg^{-1}_{\sqrt{2},-i},$$
and viceversa. \\
Thus, whatever choice we shall perform, the other Fourier-Gauss
transform is the inverse. Furthermore, since, according to the previous proposition,
$A(\sqrt{2},i)=-A(\sqrt{2},-i)$ it is immediate to conclude that $\psi(\beta)\in
Ker\left(A(\sqrt{2},i)\right)\subset\left(\mce\right)$ iff $\psi(\beta)\in Ker\left(
A(\sqrt{2},-i)\right).$
For this reason we are entitled to deal only with one of the two choices for the
Fourier-Gauss transform and, from now, $\mcg$ will stand for $\mcg_{\sqrt{2},i}$
whereas $\mcg^{-1}=\mcg_{\sqrt{2},-i}$.
}\\

To conclude the section we summarize the latter results i.e., if we start from
(\ref{KG}) and (\ref{constr}) and if we perform the Fourier-Gauss transform $\mcg$, we
end up with the following equations of motion for a $\tgbms$ massive or massless real
scalar field:

\begin{gather}\label{dyn1}
\left(-2\mcd_{\alpha(\z,\bz)}+Q_{\alpha(\z,\bz)}\right)\psi^G(\beta)=0\quad\forall
\alpha(\z,\bz)\in ST\\
\eta^{\mu\nu}\negmedspace\left(-2\mcd_{e_\mu}+Q_{e_\mu}\right)
\left(-2\mcd_{e_\nu}+Q_{e_\nu}\right)\psi^G(\beta)=\left\{
\begin{array}{ll}
0 &\;\textrm{for}\; \Delta\\
m^2\;\psi^G(\beta)&\textrm{for}\; SU(2)
\end{array}\right.,\label{dyn2}
\end{gather} 
where
$\psi^G(\beta)=\int\limits_{\mce^\prime}d\mu(\beta^\prime)\psi(\sqrt{2}\beta^\prime+i
\beta)$.\\

\remark{\label{btb}
An interesting though, to a certain extent, heuristic comment concerning (\ref{dyn2}) arises if 
we write the Klein-Gordon equation of motion for a massless scalar field $\psi$ in Minkowski spacetime  $M^4$
starting from 
$$L=\frac{1}{2}\int\limits_{M^4} d\mu(x^\rho)(-2\partial^\mu-x^\mu)\widetilde\psi(x^\rho)
(-2\partial_\mu-x_\mu)\widetilde\psi(x^\rho).$$
Here $d\mu( x^\mu )=e^{-\frac{\delta^{\mu\nu}x_\mu x_\nu}{2}}d^4x$ is the standard Gaussian measure on $\bR^4$ and
$\widetilde\psi(x^\rho)=e^{\frac{\delta^{\mu\nu}x_\mu x_\nu}{4}}\psi(x^\rho)$ is the (rescaled) real scalar field; thus $L$ is 
simply a rewriting of the usual Klein-Gordon Lagrangian and the associated equations of motion becomes:
$$\eta^{\mu\nu}(-2\partial_\mu+x_\mu)(-2\partial_\nu+x_\nu)\widetilde\psi(x^\rho)=0.$$ 
A direct inspection of this equation shows a clear resemblance with (\ref{dyn2}) which confirms the rigorously proved 
correspondence between the Poincar\'e and the $\tgbms$ massless real scalar fields (still see \cite{DMP}).
}

\section{The Lagrangian and Hamiltonian formulation of $\tgbms$ scalar field theory.}\label{cinque}
In the previous discussions we have developed the covariant approach to $\tgbms$ field 
theory exploiting the lone requirement that a free field is a suitably chosen function(al)
which transforms under a unitary and irreducible representation of the full symmetry group.
This perspective has allowed us not only to correctly identify the kinematical datum of a 
$\tgbms$ field theory but, by means of functional analysis techniques, also the dynamic of 
these fields. Nonetheless the derivation of the $\tgbms$ equations of motion (even limiting
ourselves to the real scalar case) is still unsatisfactory for two main reasons; the first 
consists of the absence of any interaction which are a key cornerstone if one wish to 
develop a complete $\tgbms$ field theory. Furthermore, from an holographic perspective, one
would like to demonstrate the existence of an holographic mapping not only for free fields
but also for the interacting ones and, in particular, we refer to gauge theories.
To this avail it is imperative to derive the $\tgbms$ equations of motion from a
variational principle and in particular we wish to
consider such a problem both in a Lagrangian  and in
an Hamiltonian framework for the "working example"of the covariant real 
(massless or massive) scalar field. 
The steps we will perform are the following: first we construct a suitable
``Lagrangian'' functional whose extremum provides (\ref{dyn1}) and (\ref{dyn2}) and then we
derive the Hamiltonian function by means of standard techniques. \\

\remark{
In order to construct the above mentioned Lagrangian, the starting point
consists of introducing a suitable space of
kinematically allowed configurations. In an infinite
dimensional setting, there are two commonly accepted and widely exploited
choices: the tangent bundle and the first jet bundle. In the latter case we should deal 
with equivalence classes of sections of an associated bundle over $\mce^\prime$.
Such a road could be pursued within our framework following the definition \ref{covdef}
for a covariant $\tgbms$ field though the characterization of a jet over the space of 
distribution over $\bS^2$ is rather tricky. 

On the other hand it is more convenient to our aims to follow the former case 
i.e. we will identify a tangent bundle over 
a suitable space of functions and an associated Lagrangian.

In the setting proper of $\tgbms$ covariant field theory we deal with, the natural
configuration space we have exploited up to now is a Fr\'echet manifold i.e. $\left(\mce\right)$ the 
space of Hida testing functionals.  It is still possible to associate to it a 
notion of tangent space: we first need to recognize that
$\left(\mce\right)$, constructed as in proposition \ref{Hidadistr}, is an abelian
ILH group and thus we are entitled to follow \cite{Omori} and to define
$T\left(\mce\right)=\bigcap\limits_p T\left(\mce\right)_p$. Furthermore, being
$\left(\mce\right)$ abelian, we may also conclude that
$T\left(\mce\right)=\left(\mce\right)\times\left(\mce\right)$.

A further option which arises and which follows more closely the usual setting of 
Poincar\'e invariant field theories consists of reminding that, according to
proposition \ref{Hidadistr}, the set of Hida testing functionals is continuously
included in $\left(L^2\right)$. Thus we can enlarge the space of kinematical
configurations to $\left(L^2\right)$. Such a choice is not only a mere convenience
if we bear in mind that both the canonical and the covariant $\tgbms$ field have been originally
introduced as function(als) on Hilbert space of square integrable functions. \\
As a matter of fact all operators involved in the construction of the previous sections, namely
$Q_{\alpha(\z,\bz)}$ and $\mcd_{\alpha(\z,\bz)}$ admits a unique continuous extension
from their natural space of definition - $\left(\mce\right)$ - to
$\left(L^2\right)$ for any $\alpha(\z,\bz)\in C^\infty(\bS^2)$ and, as outlined in 
proposition \ref{propFG}, also the Fourier-Gauss transform can be continuously extended
to a unitary operator on $\left(L^2\right)$.   

Bearing in mind these remarks we shall work in this section with $\left(L^2\right)$ whose Hilbert
structure allows us an easier identification of the Lagrangian function; we will point out in the end 
that the result holds as well in $\left(\mce\right)$.
}\\

Starting from these premises, we shall now solve the inverse ``Lagrangian'' problem i.e.
we shall start seeking for a functional $L:\left(L^2\right)\to\mathbb{R}$ whose extremum is (\ref{dyn1})
and (\ref{dyn2}). The strategy we follow consists on ignoring at the beginning
(\ref{dyn1}) requiring only that our $\tgbms$ real
massive or massless scalar field satisfies (\ref{dyn2}). To this avail, we shall 
employ a standard
technique due to Vainberg \cite{Vainberg,Bampi}. Let us remind the reader that,
given a Banach space $X$ and its dual space $X^\prime$, an operator $F:X\to
X^\prime$ is called \emph{potential} on some subset $H\subset E$ iff it exists
a functional $f$ on $X$ such that $F(x)=\nabla f(x)$ where $\nabla$ is
the gradient\footnote{We remember that an operator $F:X\to X^\prime$ is called the gradient of
a functional $f$ if $f$ admits along all directions on $X$ a Gateaux derivative
which, furthermore, must coincide with $F$.} of the functional $f$. 

Bearing in mind such a definition, the following theorem holds (we refer to 
$\S$5 in \cite{Vainberg} for the proof):\\

\teorema[Vainberg]{\label{variational}
Suppose that $X$ is a Banach space with norm $||,||$ and that 
$F:X\to X^\prime$ admits a Gateaux differential $D_hF(x)$ for all $x$ lying in a
norm induced ball $B_r(x_0)$ centered in a point $x_0\in X$ and of arbitrary but fixed 
radius $r$. Suppose also that the functional $\left(D_hF(x),h^\prime\right)$ is 
continuous in the variable $x\in B_r(x_0)$. Then $F$ is potential in $B_r(x_0)$
iff $\left(D_hF(x),h^\prime\right)=\left(D_{h^\prime}F(x),h\right)$ for all
$h,h^\prime\in X$ where $(,)$ represents the natural pairing between $X$ and
$X^\prime$.}\\

It is straightforward now to realize from the statement of this latter theorem why we
considered unsatisfactory the Fourier transform in order to formulate the $\tgbms$
equations of motion in a evolutionary form. If we wish to follow the ``traditional'' 
road of quantum field theory over Minkowski spacetime and, if we look for a formulation 
of the $\tgbms$ equations of motion (\ref{fake1}) and (\ref{fake2}) as a variational
problem, we realize that, although the operator $\eta^{\mu\nu}\mcd_{e_\mu}\mcd_{e_\nu}$
admits a Gateaux differential and $\eta^{\mu\nu}\mcd_{e_\mu}\mcd_{e_\nu}\psi(\beta)$
is continuous in the variable $\beta$, it fails to be symmetric. For a linear
operator $F$, such as the one we are dealing with, even though we should restrict ourselves
from the space $\left(\mce^\prime\right)$, where (\ref{fake1}) and (\ref{fake2}) are
naturally defined, to $\left(L^2\right)$, the symmetry condition would still imply:
\beq
\left(D_hF(x),h^\prime\right)=\left(F(h),h^\prime\right)=\left(h,F(h^\prime)
\right)=\left(D_{h^\prime}F(x),h\right),
\eeq
i.e. $F$ is self adjoint. It does not hold in our scenario since, 
exploiting lemma \ref{DQ}, one can see that 
$$\eta^{\mu\nu}\mcd_{e_\mu}\mcd_{e_\nu}(\beta)-\eta^{\mu\nu}\mcd^*_{e_\mu}
\mcd^*_{e_\nu}(\beta)=\eta^{\mu\nu}\left[\mcd_{e_\mu}Q_{e_\nu}+Q_{e_\mu}
\mcd_{e_\nu}-Q_{e_\mu}Q_{e_\nu}\right].$$
This is the first real big difference from the canonical procedure for a scalar
theory formulated in Minkowski background. Up to now, besides the complicated
techniques of white noise distribution theory, we have basically repeated
at least conceptually the same steps we would have performed in a Poincar\'e
invariant setup. At this stage, instead, we face the serious obstruction of
Vainberg theorem and this one is the main reasons why we shall adopt the Fourier-Gauss
transform $\mcg$ as in proposition \ref{propFG}. Within this framework the following
holds\\
\lemma{\label{Ldyn}
Referring to $Q$ and to $D$ as the (unique continuous) extension of the multiplication
and derivative operator from $\left(\mce\right)$ to $\left(L^2\right)$ and referring to
$e_\mu,e_\nu$ as $\left\{Y_{00}(\z,\bz),...,Y_{11}(\z,\bz)\right\}$, then 
the operator 
$$\eta^{\mu\nu}\left(Q_{e_\mu}-2\mcd_{e_\mu}
\right)\left(Q_{e_\nu}-2\mcd_{e_\nu}\right):\left(L^2\right)\to
\left(L^2\right)$$ 
is potential and the unique functional $L_{dyn}:\left(L^2
\right)\to\bR$, whose value in $\psi_0\in\left(L^2\right)$ is $L_0$, is 
\begin{gather}
L_{dyn}(\psi)=L_0+\int\limits_0^1dt\;\langle\langle\eta^{\mu\nu}\left(Q_{e_\mu}-2\mcd_{e_\mu}
\right)\left(Q_{e_\nu}-2\mcd_{e_\nu}\right)\notag\\
\label{L1}
\left(\psi_0+t(\psi-\psi_0)\right),\psi-\psi_0\rangle\rangle_{\left(L^2\right)},
\end{gather}
where $\langle\langle,\rangle\rangle_{\left(L^2\right)}$ is the internal product
\footnote{We adopt the symbol $\langle\langle,\rangle\rangle$ which stands for the 
natural pairing between $\left(\mce\right)$ and $\left(\mce^\prime\right)$ because it
is subject to the compatibility condition (\ref{compat2}) according to which it
coincides with the internal product on $\left(L^2\right)$ when we evaluate 
$\langle\langle\phi,\phi^\prime\rangle\rangle$ with $\phi\in\left(\mce\right)$ and 
$\phi^\prime\in\left(L^2\right)$.} on
$\left(L^2\right)$.}\\
 
\emph{Proof.}
Identifying the Hilbert space $\left(L^2\right)$ with its dual by means of the
Riesz theorem, the operator $A=\eta^{\mu\nu}\left(Q_{e_\mu}-2\mcd_{e_\mu}
\right)\left(Q_{e_\nu}-2\mcd_{e_\nu}\right)$ is a map
from $\left(L^2\right)$ to $\left(L^2\right)^\prime$. It admits a continuous
Gateaux derivative\footnote{The definition Gateaux derivative on a functional from
$\left(L^2\right)$ to $\bR$ is a straightforward adaptation of definition
\ref{deriv}. For this reason, we feel that, for the economy of the paper, it is
useless to introduce an additional symbol and we will use also in this case 
$\mcd$. The associated pedex will univocally distinguish between the different cases.} for all $\psi\in\left(L^2\right)$ and along all directions 
$\psi^\prime\in\left(L^2\right)$ since, per definition (\ref{Gateaux}) and,
being $A$ linear, 
$$\mcd_{\psi^\prime}\left[\eta^{\mu\nu}\left(Q_{e_\mu}-2\mcd_{e_\mu}
\right)\left(Q_{e_\nu}-2\mcd_{e_\nu}\right)\right]\psi=
\eta^{\mu\nu}\left(Q_{e_\mu}-2\mcd_{e_\mu}
\right)\left(Q_{e_\nu}-2\mcd_{e_\nu}\right)\psi^\prime.$$
Furthermore, on $\left(L^2\right)$ the operator under analysis is according to
proposition \ref{propFG} selfadjoint thus symmetric. The
hypotheses of Vainberg theorem are met and
$\eta^{\mu\nu}\left(Q_{e_\mu}-2\mcd_{e_\mu}\right)\left(Q_{e_\nu}-2\mcd_{e_\nu}
\right)$ is potential. Uniqueness of the functional - i.e. (\ref{L1}) - 
whose gradient satisfies the equation (\ref{dyn1}) is now a direct consequence
of Vainberg theorem which grants us that
$$\mcd_{\psi^\prime}L(\psi)=\langle\langle\eta^{\mu\nu}\left(Q_{e_\mu}
-2\mcd_{e_\mu}\right)\left(Q_{e_\nu}-2\mcd_{e_\nu}\right)\psi,\psi^\prime\rangle
\rangle_{\left(L^2\right)}.$$
Thus for any $\psi$ in a ball
$D=\left\{\psi\in\left(L^2\right)\:\left|\right.\:\langle\langle\psi-\psi_0
\rangle\rangle_{\left(L^2\right)}<r\right\}$ centered in $\psi_0$ and of fixed
radius $r$, and for any $t\in\left[0,1\right]$ the last equality translates as
\begin{gather*}\frac{d}{dt} L\left(\psi_0+t(\psi-\psi_0)\right)=\\
=\langle\langle\eta^{\mu\nu}
\left(Q_{e_\mu}-2\mcd_{e_\nu}\right)\left(Q_{e_\nu}-2\mcd_{e_\nu}\right)\left(
\psi_0+t(\psi-\psi_0)\right),\psi-\psi_0\rangle\rangle.
\end{gather*}
An integration in the $t$ variable shows that (\ref{L1}) is the unique
functional whose gradient is our equation $\eta^{\mu\nu}\left(Q_{e_\mu}
-2\mcd_{e_\mu}\right)\left(Q_{e_\nu}-2\mcd_{e_\nu}\right)\psi(\beta)=0$. $\Box$\\

\remark{
Setting the initial condition as $\psi_0=0$, $L_0=0$ and adding the mass term 
whenever we wish to deal with a massive $\tgbms$ real scalar field, then 
(\ref{L1}) becomes:
\begin{gather}\label{LD1}
L_{KG}(\psi)=\frac{1}{2}\langle\langle\left[\eta^{\mu\nu}\left(Q_{e_\mu}
-2\mcd_{e_\mu}\right)\left(Q_{e_\nu}-2\mcd_{e_\nu}\right)+m^2\right]\psi,\psi
\rangle\rangle_{\left(L^2\right)}=\notag\\
\frac{1}{2}\int\limits_{\mce^\prime}d\mu(\beta)\eta^{\mu\nu}\left[-4\mcd_{e_\mu}
\psi(\beta)\mcd_{e_\nu}\psi(\beta)+\right.\notag\\
\left.(\beta,e_\mu)(\beta,e_\nu)\psi^2(\beta)+
4(\beta,e_\mu)\psi(\beta)\mcd_{e_\nu}\psi(\beta)+m^2\psi^2(\beta)\right],
\end{gather}
where in the last equality we have exploited the definition of multiplication
operator and lemma \ref{DQ} whereas $(\beta,e_\mu)$ still stands for the 
canonical pairing between $\mce^\prime$ and $C^\infty(\bS^2)$.

We will refer to this term as the \emph{Klein-Gordon part of the Lagrangian} for 
the $\tgbms$ massive or massless scalar field. It is also imperative to
underline that the above methods can be fully applied also to non scalar
$\tgbms$ field without any substantial modifications in the reasoning and in the
demonstrations. This still confirms that we are working with a specific field
only for the sake of simplicity and of clarity nonetheless without losing in
generality.
}\\

We face now the last obstacle i.e. we need also to implement (\ref{dyn2}). A
direct inspection shows that (\ref{dyn2}) is a family of constraints on the 
covariant fields and thus we seek to implement it in terms of Lagrange multipliers:\\

\proposizione{
The functional $L:\left(L^2\right)\to\bR$ whose associated Euler 
equations are (\ref{dyn1}) and (\ref{dyn2}) is
\beq\label{LD2}
L(\psi,\lambda_i)=L_{KG}(\psi)+\sum\limits_i\int\limits_{\mce^\prime}d\mu(\beta)
\frac{\lambda_i}{2}(\beta)\left[\left(-2\mcd_{e_i}+Q_{e_i}\right)\psi(\beta)\right]^2,
\eeq
where $L_{dyn}$ is (\ref{LD1}), $e_i=\left\{Y_{lm}(\z,\bz)\right\}_{l>1}$ 
whereas $\lambda_i(\beta)\in\left(L^2\right)$ are suitable Lagrange multipliers.
Furthermore, as previously, the operator $Q$ refers to the unique 
continuous extension to $\left(L^2\right)$ of the corresponding Gateaux 
derivative and multiplication operator on $\left(\mce\right)$.}\\
\emph{Proof.}
The first step in the demonstration consists of showing that an element
$\psi(\beta)\in\left(L^2\right)$ satisfies $\left(-2\mcd_{\alpha(\z,\bz)}+Q_{
\alpha(\z,\bz)}\right)\psi(\beta)=0$ for all $\alpha(\z,\bz)\in ST$ iff 
$\left(-2\mcd_{e_i}+Q_{e_i}\right)\psi(\beta)=0$ for each $e_i$. This statement
straightforwardly holds since, according to theorem \ref{decompos}, 
$ST$ is the closed set of real linear combinations of the real spherical
harmonics with $l>1$ and since the operator $-2\mcd+Q$ seen as a map from
$C^\infty(\bS^2)\times\left(L^2\right)\to\left(L^2\right)$ mapping the pair 
$(\alpha(\z,\bz),\psi(\beta))$ into $\left(-2\mcd_{\alpha(\z,\bz)}+Q_{\alpha(\z,
\bz)}\right)\psi(\beta)$ is linear in the first argument.

The remaining part of the proof will be structured as follows: we will calculate the variation with 
respect to  $\psi$ of a generic functional 
$$L(\psi)=\int\limits_{\mce^\prime}d\mu(\beta)\mathcal{L}
(\beta,\psi(\beta),\mcd_{\beta^\prime}\psi(\beta)),$$ 
where $\mathcal{L}:\left(L^2\right)\to\left(L^2\right)$ is a ``density'' 
depending both on the fields and on its derivative along any 
direction. The final result will be the ``Euler-Lagrange'' equation associated 
to a functional defined on a space endowed with a Gaussian measure. 
Eventually we will apply the result to (\ref{LD2}).

Let us thus perform the following variation: pick any
$\phi(\beta)\in\left(L^2\right)$, then
$$\langle\langle\frac{\delta L}{\delta \psi},\phi(\beta)\rangle\rangle=
\lim\limits_{\epsilon\to 0}\frac{1}{\epsilon}\left[L(\beta,\psi+\epsilon\phi,
\mcd_{\beta^\prime}(\psi+\epsilon\phi))-L(\beta,\psi,\mcd_{\beta^\prime}\psi)
\right]=$$
$$=\int\limits_{\mce^\prime}d\mu(\beta)\;\mcd_{\psi}\mathcal{L}(\beta,\psi,
\mcd_{\beta^\prime}\psi)\phi(\beta)+\int\limits_{\mce^\prime}d\mu(\beta)\;
\mcd_{\mcd_{\beta^\prime}\psi}\mathcal{L}(\beta,\psi,\mcd_{\beta}\psi)
\mcd_{\beta^\prime}\phi(\beta).$$
The second element in the right hand side of the last equality can be written in
the more convenient form
$$\langle\langle\mcd_{\mcd_{\beta^\prime}\psi}\mathcal{L}(\beta,\psi,
\mcd_{\beta}\psi),\mcd_{\beta^\prime}\phi(\beta)\rangle\rangle=
\langle\langle\mcd_{\beta^\prime}^*\left[\mcd_{\mcd_{\beta^\prime}\psi}\mathcal{L}
(\beta,\psi,\mcd_{\beta}\psi)\right],\phi(\beta)\rangle\rangle=$$
$$=\langle\langle\left(-\mcd_{\beta^\prime}+Q_{\beta^\prime}\right)
\mcd_{\mcd_{\beta^\prime}\psi}\mathcal{L}(\beta,\psi,\mcd_{\beta}\psi),\phi(
\beta)\rangle\rangle,$$
where $\mcd_{\beta^\prime}^*$ is the adjoint derivative operator and where we 
have exploited the relation $\mcd_{\beta^\prime}^*+\mcd_{\beta^\prime}=Q_{\beta^
\prime}$ as in lemma \ref{DQ}.

Thus in order for the variation of (\ref{LD2}) to vanish for any choice of
$\phi(\beta)$, we end up with the following Euler-Lagrange equation:
\beq\label{EL}
\mcd_{\psi}\mathcal{L}(\beta,\psi,\mcd_{\beta^\prime}\psi)-\left(\mcd_{
\beta^\prime}-Q_{\beta^\prime}\right)\mcd_{\mcd_{\beta^\prime}\psi}\mathcal{L}(
\beta,\psi,\mcd_{\beta}\psi)=0,
\eeq
where the term with the multiplication operator is a feature typical due to the 
presence of a Gaussian measure $\mu$ on $\mce^\prime$. 

This formula can be straightforwardly extended when, as in the scenario under
consideration, the functional depends upon more than one field. Thus 
a straightforward application of (\ref{EL}) in (\ref{LD2}) shows that the
variation of functional for the scalar $\tgbms$ field with respect to the 
Lagrange multipliers provides that $\left(-2\mcd_{e_i}+Q_{e_i}\right)\psi(\beta)
=0$ for all $e_i$
whereas a variation with respect to $\psi$ provides, once the constraints are
imposed, equation (\ref{KG}). $\Box$\\

\remark{
In the wake of the above proposition, it is natural and at same time imperative to wonder ourselves
whether we are titled to really refer to (\ref{LD2}) as the Lagrangian of our system. As a matter of fact 
a direct inspection of (\ref{LD2}) shows that the functional under analysis can also be interpreted as 
map $L:T\left(L^2\right)\to\bR$ which associates to the pair $(\psi,\mcd_{e_0}\psi)\in T\left(L^2\right)\equiv
\left(L^2\right)\times\left(L^2\right)$ the expression 
(\ref{LD2}). Here $\mcd_{e_0}$ represents the natural counterpart for the time derivative of a
covariant field in Minkowski spacetime.

Furthermore, bearing in mind that the Gateaux derivative along any direction on $\mce^\prime$ is a
continuous map from $\left(\mce\right)$ into itself, it is immediate to realize that the above 
interpretation for (\ref{LD2}) holds also if we refer to Hida testing functionals. Thus
$L(\psi,\lambda_i)$
can also be taught as the Lagrangian for the $\tgbms$ scalar field on $T\left(\mce\right)$. 
}\\

Having solved the inverse Lagrangian problem, we are now apparently in position to
formulate the free $\tgbms$ field theory in an Hamiltonian framework. While the 
Lagrangian analysis is best performed in the tangent space of a suitably chosen
configuration space, the Hamiltonian counterpart is naturally developed with the
tools proper of symplectic geometry. Since it will also play a fundamental role
in the forthcoming analysis we will first choose a symplectic space and the
natural obvious choice is the cotangent bundle over our configuration space.
If one wishes to work directly with the the space of Hida testing functional,
it is natural to resort again to the identification of $\left(\mce\right)$ with
an abelian ILH which leads to identify
$T\left(\mce\right)=\left(\mce\right)\times\left(\mce\right)$ and, per duality,
$T^*\left(\mce\right)=\left(\mce\right)\times\left(\mce^\prime\right)$. On the
opposite we will choose as configuration space $\left(L^2\right)$ which, bearing
in mind that $T\left(L^2\right)=\left(L^2\right)\times\left(L^2\right)$, allows
us to identify by Riesz theorem
$T^*\left(L^2\right)=\left(L^2\right)\times\left(L^2\right)$. 

As for the Lagrangian counterpart we will focus first on the cotangent bundle with an
Hilbert structure remarking in the end that all the results can be applied also in the former 
case without any significant modification.\\

\definizione{\label{phasesp}
We call $\Gamma$, the Cartesian product $\left(L^2\right)\times\left(L^2\right)$
the \emph{phase space} of a $\tgbms$ real (massive or massless) scalar field
associated to the configuration space $\left(L^2\right)$. If the latter is
chosen as $\left(\mce\right)$ then
$\Gamma=\left(\mce\right)\times\left(\mce^\prime\right)$.
}\\

\noindent Following nomenclatures of \cite{Chernoff,Schmid}:\\

\proposizione{
The vector space $\Gamma$ endowed with the continuous bilinear map (with respect
to the product topology) $\Omega:\Gamma\times\Gamma\to\bR$
\beq
\Omega\left((\psi_1,\Psi_1),(\psi_2,\Psi_2)\right)=\langle\langle\Psi_2,\psi_1
\rangle\rangle-\langle\langle\Psi_1,\psi_2\rangle\rangle
\eeq 
is a symplectic vector space. Here $\langle\langle,\rangle\rangle$ is either the
canonical pairing between $\left(\mce\right)$ and $\left(\mce^\prime\right)$ or
the internal product on $\left(L^2\right)$ depending on the chosen phase space.}\\

\emph{Proof.}
We need only to show that $\Omega$ is a weakly non degenerate skew symmetric 
bilinear form. Independently from the two possible cases in the hypotheses, 
skew-symmetry is trivially verified whereas, in order to show the weak non 
degenerateness of $\Omega$, we need to 
show that, calling $\nu_1=(\psi_1,\Psi_1)$ and $\nu_2=(\psi_2,\Psi_2)$, then
$\Omega(\nu_1,\nu_2)=0$ for any $\nu_2\in\Gamma$ implies $\nu_1=0$. Choose $\nu_2=(0,\Psi_2)$; then
$\Omega(\nu_2,\nu_2)=\langle\langle\Psi_2,\psi_1\rangle\rangle=0$ for any choice
of $\Psi_2\in\left(\mce^\prime\right)$. This is possible iff $\psi_1=0$. Similarly choose now $\nu_2=(\psi_2,0)$; 
accordingly $\Omega(\nu_2,\nu_1)=\langle\langle\Psi_1,\psi_2\rangle\rangle=0$ for any choice 
of $\psi_2\in\left(\mce\right)$. This is achievable only if $\Psi_1=0$. 
$\Box$\\

\remark{
It is interesting to pinpoint that, if we resort to work on nuclear spaces 
such as $\left(\mce\right)$, we are constrained to deal only with Fr\'echet 
structures which, thus, forbids us to select
a strongly non degenerate symplectic space which is the natural structure in 
finite dimensional dynamical systems. On the opposite, if we choose to work on 
the Hilbert spaces ,such as $\left(L^2\right)$, it is straightforward to 
realize, still thanks to Riesz theorem, that $\Omega^b:\Gamma\to\Gamma^*$,
mapping $\nu\in\Gamma$ into the linear operator $\Omega(\nu):\Gamma\to\bR$ is 
an isomorphism and thus the symplectic form is \emph{strongly non degenerate}. 

Thus from now we will consider only the symplectic phase space
$\left(T^*\left(L^2\right),\Omega\right)$. 
}\\

Bearing in mind the above comments one immediately realize that the construction
of the Hamiltonian function is not a straightforward calculation since
(\ref{LD2}) is a singular Lagrangian. Thus we need to resort to the theory of
constraints and in particular to the algorithm developed by Gotay, Nester and
Hinds in \cite{Gotay,Gotay2} which is a geometrization and a generalization of 
the canonical Dirac-Bergman theory. In particular here we will adapt to our
Hilbert configuration manifold the analysis of a Lagrangian system with Lagrange
multipliers performed in \cite{Carinena,Martinez} for finite dimensional
configuration spaces. 

Since the constraints (\ref{dyn2}) are globally defined on $T\left(L^2\right)$,
the first natural step consists of promoting the multipliers in (\ref{LD2}) to
dynamical variable thus switching from $T\left(L^2\right)$ to $TP\equiv T\left[\left(L^2
\right)\times\left(L^2\right)^N\right]$ where $\left(L^2\right)^N$ means that we
consider as many copies of $\left(L^2\right)$ as the number of needed Lagrange
multipliers. In the case under consideration this is equal to the number of
spherical harmonics with $l>1$. Local coordinates on $P$ are given by
$(\psi,\mcd_{e_0}\psi,\lambda_i,\mcd_{e_0}\lambda_i)$ where now $e_0$
stands for the $l=0$ spherical harmonic; it plays the role of the time direction in a Poincar\'e
invariant theory. Reading (\ref{dyn2}) as a map $L:TP\to\bR$, we can
introduce the fiber derivative $\mcf L:TP\to T^*P$ such that 
$$\langle\langle\psi^\prime,\mcf L(\psi)\rangle\rangle_{\left(L^2\right)}=\left.
\frac{d}{dt}L(\psi+t\psi^\prime)\right|_{t=0}=\mcd_{\psi^\prime}L(\psi)=
\langle\langle\mcd L,\psi^\prime\rangle\rangle_{\left(L^2\right)}.$$
In local coordinates such a transformation becomes
\beq
\mcf L(\psi,\mcd_{e_0}\psi,\lambda_i,\mcd_{e_0}\lambda_i)=\left(\psi,\mcd_{\mcd\psi}L,
\lambda_i,0\right)=
\left(\psi,4\mcd_{e_0}\psi-2Q_{e_0}\psi,\lambda_i,0\right).\label{FL}
\eeq
The above equality simply restates that the Lagrangian function is not
hyperregular and thus the fiber derivative is not a diffeomorphism. Consequently we 
are obstructed to introduce the Hamiltonian as $H=E\circ \mcf L^{-1}$ where $E$ represents
the energy function
$$E=\langle\langle\psi,\mcf L(\psi)\rangle\rangle_{\left(L^2\right)}-L(\psi).$$ 
On the opposite
we may still construct it implicitly on the image of $\mcf L(TP)$ as $H\circ\mcf L=E$
which is a reasonable definition iff for any two points $p,p^\prime\in TP$ such that 
$\mcf L(p)=\mcf L(p^\prime)$ then $E(p)=E(p^\prime)$.

As discussed mainly in \cite{Gotay2},  such last condition is satisfied if
the Lagrangian under analysis is \emph{almost regular} 
i.e. $\mcf L$ is a submersion onto $T^*P$ and, for any $p\in TP$, the fibers 
$(\mcf L)^{-1}\left\{\mcf L(p)\right\}$ are connected submanifolds of $TP$. \\ 

\proposizione{
The functional (\ref{LD2}) is an almost regular Lagrangian.}\\

\emph{Proof.}
The demonstration is divided in two parts: first we  show that $\mcf L(TP)$ is a 
submersion and than we prove that the fibers are connected submanifolds.

In order to deal with the first assertion we exploit proposition 2.2 in chapter II $\S$2 of
\cite{Lang} according to which a class $C^p$ ($p\geq 0$) morphism $f$ between two manifolds
of class $C^p$ - $X,Y$ - modelled over Banach spaces is a submersion at $x\in X$ iff it exists a 
chart $(U,\varphi)$ at $x$ and a second chart $(V, \phi)$ at $f(x)\in Y$ such that 
$\mcd f_{V,U}(\varphi(x))$ is surjective and the kernel splits

In the hypotheses of this proposition both $TP$ and $T^*P$ are Hilbert spaces which can be
identified exploiting Riesz theorem. Thus, choosing any 
chart centerd at a point $p\in TP$,  
a direct inspection of (\ref{FL}) shows either that the fiber derivative is a surjection on its image 
either that the kernel of $\mcf L$ is the set of real linear
combinations of vectors $\left(0,0,0,\mcd\lambda_i\right)$. Thus $Ker\left(\mcf L(TP)\right)$ is
isomorphic to $\left(L^2\right)^N$ and $TP=Ker(\mcf L) +M_1$ where $M_1=\left(L^2\right)\times 
\left(L^2\right)\times\left(L^2\right)^N$ with $M_1\cap Ker(\mcf L)=\left\{0\right\}$. This latter
decomposition induces a natural map
from $TP$ into the Cartesian product $Ker(\mcf L)\times M_1$ which is a (toplinear) isomorphism and 
thus the kernel splits.

Concerning the second part of the demonstration, consider any point $q\in T^*(P)$ such that $q=\mcf L(\bar p)$ 
with $\bar p\in TP$. Pick any two points - say $p_1, p_2$ lying in $\mcf L^{-1}(q)$. Referring to $\tau: TP\to P$ as the
tangent bundle projection map and to $\pi:T^*P\to P$ as the cotangent bundle counterpart, we may conclude from
the compatibility condition $\pi\circ\mcf L=\tau$ that $\tau(p_1)=\tau(p_2)$ i.e.
$(\psi_1,\lambda_{i1})=(\psi_2,\lambda_{i2})$. \\
To conclude the demonstration it is sufficient 
now to exploit (\ref{FL}) and the hypothesis $\mcf L(p_1)=\mcf L(p_2)$ according to which 
$$\left(\psi_1,4(\mcd_{e_0}\psi)_1-2Q_{e_0}\psi_1,\lambda_{i1},0\right)=\left(\psi_2,4(\mcd_{e_0}
\psi)_2-2Q_{e_0}\psi_2,\lambda_{i2},0\right).$$

It implies that the two points $p_1,p_2$ differ at most for an element in $Ker(\mcf L)$; thus the
fibers are connected submanifolds. $\Box$\\

As a consequence of this last theorem, we know that the energy function in constant along the fibers of $\mcf L$ and thus it 
induces on the manifold $M_1$ a well defined Hamiltonian function as
\begin{gather}
H(\psi,\lambda_i,\Pi)=\frac{1}{2}\int\limits_{\mce^\prime}d\mu(\beta)\left[\Pi^2(\beta)-\sum\limits_{k=1}^3
\left[Q_{e_k}\psi(\beta)-2\mcd_{e_k}\psi(\beta)\right]^2-m^2\psi^2(\beta)\right]+\notag\\
-\frac{1}{2}\sum\limits_i\int\limits_{\mce^\prime}d\mu(\beta)
\lambda_i(\beta)\left[\left(-2\mcd_{e_i}+Q_{e_i}\right)\psi(\beta)\right]^2,
\end{gather}
where $\Pi=-2\mcd_{e_0}\psi(\beta)+Q_{e_0}\psi(\beta)$ is the conjugate momentum whereas $e_k$ are the three spherical 
harmonics direction in $C^\infty(\bS^2)$ with $l=1$. 

\section{Comments and conclusions}
The overall results of this paper could simply be summarized with the set
phrase ``the circle has been closed''.  Starting from \cite{Arcioni}, it
was realized that the infinite dimensional nature of the supertranslations and
of the supermomenta forces us to  deal with $\tgbms$ fields being functionals instead of 
the canonical functions proper of a Poincar\'e invariant theory over Minkowski spacetime
or more generally of a quantum field theory over a curved background.

Consequently it appeared that only the purely group theoretical Wigner 
programme could shed some light
on the kinematically and dynamically allowed configurations for a BMS invariant 
field theory living at future (or past) null infinity; the paradigm of equations 
of motion as an extremum out of a variational principle was thus a priori discarded.

Such an obstruction was previously gotten around exploiting the rigorous means
of algebraic quantum field theory out of which some ``holographic theorems" were proved.  
In this paper we wished to overcome the above deficiency ad we managed to associate to
a scalar $\tgbms$ field theory a genuine Hamiltonian system. To achieve such a goal 
we followed the path to rigorously define and analyze the covariant formulation of 
a $\tgbms$ invariant theory. Within this framework each field arises as an element 
in a suitably constructed space of Hida testing functionals or, more generally, in its univocally
associated Gelfand triplet. 

This novel point of view lead us to a twofold result: as a first step we casted the equations of motion 
for a $\tgbms$ field as suitable operators acting on the above mentioned space of Hida testing functionals.
Afterwards, by a continuous extension, to its Hilbert space completion, we have shown that each 
equation of motion for a real massive or massless scalar field could be interpreted as the 
Euler-Lagrange equation of a  suitable functional.\\
Alas, such a Lagrangian turned out not be hyperregular and thus the fiber derivative from the tangent
to the cotangent space over the set of kinematical configuration is not a diffeomorphism. Exploiting 
the geometric description of the constraint algorithm originally due to Nester and Goaty for presymplectic Lagrangian
manifolds, we have nonetheless manged to show that on a suitable connected submanifold of the symplectic
cotangent bundle, we could identify an Hamiltonian function.

Compared to Ashtekar and Streubel result, a direct inspection shows that, since our analysis starts from
an intrinsic definition of a $\tgbms$ field theory, it enlightens the contribution of the pure supertranslational 
component of $C^\infty(\bS^2)$ which appeared to be partially neglected in \cite{Ashtekar}. Consequently we confirm 
the conclusions sketched already in \cite{Arcioni2} according to which the  result in \cite{Ashtekar} 
encompasses mainly the datum from what we referred to as the Klein-Gordon component of the 
$\tgbms$ dynamical system.

From a future perspective, one could claim that, on a physical ground, the results achieved put us into the
position to discuss without further ado if an holographic correspondence between bulk and boundary (Yang-Mills)
gauge theories really exists in an asymptotically flat spacetime. As already mentioned in the introduction the next
direct step after our analysis starts from the results of section 4 and 5 leading to the development of symplectic 
techniques out of which we may construct a $\tgbms$ interacting field theory.

From a mere holographic point of view, although it was more an underlying motivation for the whole
line of research rather than for this specific paper, we can nonetheless comment that 
we have now better clarified, from the functional analytic point of view, 
the existence of the bulk to boundary correspondence for massless real scalar fields proved in \cite{DMP}. 
In particular remark \ref{btb} outlined that the relevant operators, describing the dynamic of the field theory 
both in a flat background and at null infinity, are ultimately the same.
As a side remark, one could also hope that such a line of thinking could shed some light on the problem, mentioned
in the introduction, to construct a full holographic correspondence for massive free field. 
Within the ``functional perspective'' there is no apparent obstruction to relate massive fields on Minkowski and on
its conformal boundary and thus the obstruction lies in developing a concrete geometrical way to project the data from the 
bulk to null infinity itself.

From a pure mathematical point of view it appears that the realization of BMS field theory as a dynamical system 
can be coherently and fully described in terms of white noise analysis. The only minor obstruction to the date consists in
the  ``tangent bundle'' approach. In a finite dimensional counterpart, it is common to formulate classical field theory in 
terms of jet bundles which allow to treat on the same ground time and spatial derivatives. Such a problem clearly
arises also in a $\tgbms$ framework where one wishes to encompass in a unique setting all the Gateaux derivatives 
along $\mce^\prime$-directions. Unfortunately, as outlined in section 4, covariant $\tgbms$ fields are maps
from $\mce^\prime$ into a suitable target space and the former is not a priori a Fr\'echet manifold but simply a locally
convex topological space. Thus it appears to be rather difficult, or at least unknown to us,  how
to coherently introduce, within the $\tgbms$ framework, the notion of (first) jet bundle; the most promising 
road within this direction lies in a sheaf theoretical formulation of the Hamiltonian theory though
it would possibly forbid us to deal with global issues addressing only the local ones. 
We will analyze in detail such a problem in a future paper.

\subsection*{Acknowledgements}
The author is in great debt with V. Moretti, M. Carfora and O. Maj for several 
usefull discussions during the realization of this manuscript. The work 
has been supported partially by a grant from the Department of Theoretical and Nuclear Physics 
of Pavia University and partially by a grant from GNFM-INdAM (Istituto Nazionale Di Alta Matematica) under the project 
\emph{``Olografia e spazitempo asintoticamente piatti: un approccio rigoroso''}.


\begin{thebibliography}{100}
\bibitem{'thooft}
  G.~'t Hooft,
  \emph{``Dimensional reduction in quantum gravity,''}
  arXiv:gr-qc/9310026,

\bibitem{Aharony}
  O.~Aharony, S.~S.~Gubser, J.~M.~Maldacena, H.~Ooguri and Y.~Oz,
  \emph{``Large N field theories, string theory and gravity,''}
  Phys.\ Rept.\  {\bf 323} (2000) 183
  [arXiv:hep-th/9905111],

\bibitem{Rehren}
  K.~H.~Rehren,
  \emph{``Algebraic holography,''}
  Annales Henri Poincare {\bf 1} (2000) 607
  [arXiv:hep-th/9905179],
  
\bibitem{Duetsch}
  M.~Duetsch and K.~H.~Rehren,
  \emph{``Generalized Free Fields And The Ads-Cft Correspondence,''}
  Annales Henri Poincare {\bf 4} (2003) 613
  [arXiv:math-ph/0209035],

\bibitem{Albeverio}
S.~Albeverio, A.~Hahn and A.~N.~Sengupta
\emph{``Rigorous Feynmann path integrals with applications to quantum theory,
gauge fields and topological invariants''} in \emph{Stochastic analysis and mathematical physics} (2004) 
World Scientific,

\bibitem{Kuo} H.-H. ~Kuo, 
\emph{``White noise distribution theory''} (1996) CRC Press,
  
\bibitem{Dappiaggi}
C.~Dappiaggi,
\emph{``BMS field theory and holography in asymptotically flat space-times,''}
JHEP {\bf 0411} (2004) 011
[arXiv:hep-th/0410026],
  
\bibitem{deboer}
J.~de Boer, S.~N.~Soldukhin,
\emph{``A holographic reduction of Minkowski spacetime''}
Nucl. Phys. B {\bf 665} (2003) 545
[arXiv:hep-th/0303006],

\bibitem{Mann}
  R.~B.~Mann and D.~Marolf,
  \emph{``Holographic renormalization of asymptotically flat spacetimes,''}
  Class.\ Quant.\ Grav.\  {\bf 23} (2006) 2927
  [arXiv:hep-th/0511096],
  
\bibitem{Arcioni}
G.~Arcioni and C.~Dappiaggi,
\emph{``Exploring the holographic principle in asymptotically flat spacetimes  via
the BMS group,''}
Nucl.\ Phys.\ B {\bf 674} (2003) 553
[arXiv:hep-th/0306142].

\bibitem{Arcioni2}
G.~Arcioni and C.~Dappiaggi,
\emph{``Holography in asymptotically flat spacetimes and the BMS group''}
Class. Quant. Grav. {\bf 21} (2004) 5655,

\bibitem{Moretti}
  V.~Moretti,
  \emph{ ``Uniqueness theorem for BMS-invariant states of scalar QFT on the null
   boundary of asymptotically flat spacetimes and bulk-boundary observable
  algebra correspondence,''}
  arXiv:gr-qc/0512049, to appear on Comm. Math. Phys., 
    
\bibitem{DMP} 
C.~Dappiaggi, V.~Moretti and N.~Pinamonti,
\emph{``Rigorous steps towards holography in asymptotically flat
spacetimes,''} arXiv:gr-qc/0506069  Rev. Math. Phys. {\bf 18} (2006) 349,  

\bibitem{Weinstein}
A.~Weinstein,
\emph{``A Universal Phase Space For Particles In Yang-Mills Field,''}
Lett.\ Math.\ Phys.\  {\bf 2} (1978) 417,

\bibitem{Guillemin}
V.P. Guillemin, S. Sternberg
\emph{``Symplectic techniques in physics''} (1984) Cambridge University Press,

\bibitem{Landsman}
N.P. Landsman: \emph{``Mathematical topics between classical and quantum
mechanics''} (1998) Springer,

\bibitem{Ashtekar}
  A.~Ashtekar and M.~Streubel,
  \emph{``Symplectic Geometry Of Radiative Modes And Conserved Quantities At Null
  Infinity,''}
  Proc.\ Roy.\ Soc.\ Lond.\ A {\bf 376} (1981) 585,

\bibitem{Chernoff}
P.R. ~Chernoff and J.E. ~Marsden,
\emph{``Properties of Infinite Dimensional Hamiltonian Systems,''}
Springer-Verlag (1974),

\bibitem{Mc1} P.J. McCarthy, 
\emph{``The Bondi-Metzner-Sachs in the nuclear topology''} 
Proc. R. Soc. London {\bf A343} (1975) 489,
  
\bibitem{Wald} 
R.~M.~Wald, 
{\it ``General Relativity''},
Chicago University Press, Chicago (1984),

\bibitem{Penrose} R. Penrose, 
\emph{``Asymptotic Properties of Space and Time''} 
Phys. Rev. Lett. {\bf 10} (1963) 66,

\bibitem{Penrose2} R. Penrose, in: A.O. Barut (Ed.), \emph{``Group Theory in 
Non-Linear Problems''}, Reidel, Dordrecht (1974), p. 97 chapter 1,

\bibitem{Geroch} R. Geroch, in: P. Esposito, L. Witten (Eds.) \emph{``Asymptotic 
Structure of Spacetime''}, Plenum, New York (1977),

\bibitem{Gelfand} I. M. Gel'fand {\em et al.}, \emph{``Generalized functions:
Integral Geometry and Representation Theory, Vol. 5''} (1966) Acadmic Press,

\bibitem{Hida} T. ~Hida, H.-H. ~Kuo, N. ~Obata, 
\emph{``Transformations for white noise functionals''} 
J. Funct. Anal. {\bf 111} (1993) 259,

\bibitem{Hida2} T. ~Hida, H.-H. ~Kuo, J. ~Potthoff, L. ~Streit
\emph{``White Noise: an infinite dimensional calculus''} (1993) Kluwer Academics
Publishers,

\bibitem{Group} A.O. Barut, R. Raczka: \emph{``Theory of group representation and
applications''} World Scientific 2ed (1986),

\bibitem{Gelfand2} I. M. Gel'fand {\em et al.}, \emph{``Generalized functions:
Application of Harmonic analysis, Vol. 4''} (1966) Acadmic Press,

\bibitem{Omori} H.~Omori,
\emph{``Infinite dimensional Lie Groups''} (1974) Springer-Verlag,

\bibitem{Simms} D.J. Simms: \emph{``Lie groups and quantum mechanics''}
Springer-Verlag (1968),

\bibitem{Mackey} G. W. Mackey: \emph{``Unitary Group Representations in Physics, Probability and Number Theory''}
Addison-Wesley Publishing (1989),

\bibitem{LLedo} F.~LLedo, 
\emph{``Massless relativistic wave equations and quantum field theory''}
Ann.\ Henri\ Poincar\'e {\bf 5} (2004) 607,

\bibitem{Piard}
  A.~Piard,
  \emph{``Unitary Representations Of Semidirect Product Groups With Infinite
  Dimensional Abelian Normal Subgroup,''}
  Rept.\ Math.\ Phys.\  {\bf 11} (1977) 259,

\bibitem{Kubo} I. ~Kubo, H.-H. ~Kuo
\emph{``Finite dimensional Hida distributions''}
J. Funct. Anal. {\bf 128} (1995) 1,

\bibitem{Lee} J.Y.~Lee \emph{``Integral transforms of analytic functions on abstract 
Wiener spaces´´}
J. Funct. Anal. {\bf 47} (1982) 153,

\bibitem{Vainberg}
M.~M.~Vainberg, 
\emph{``Variational Methods for the Study of Nonlinear Operators''}
Holden-Day, Inc. (1964),

\bibitem{Bampi}
F.~Bampi and A.~Morro
\emph{``The inverse problem of the calculus of variations applied to continuum
physics''}
J. Math. Phys. {\bf 23} (1982) 2312,

\bibitem{Schmid}
R.~Schmid,
\emph{``Infinite dimensional Hamiltonian Systems''}
Bibliopolis (1987),

\bibitem{Gotay}
M.~Gotay, J.M.~Nester and G.~Hinds
\emph{``Presymplectic manifolds nad the Dirac-Bergmann theory of constraints''}
J. Math. Phys {\bf 19} (1978) 2388,

\bibitem{Gotay2}
M.~Gotay and J.M.~Nester
\emph{``Presymplectic Lagrangian systems I: the constraint algorithm and the
equivalence theorem''}
Ann. Inst. Henri Poincar\`e - Sec. A {\bf 30} (1979) 129,

\bibitem{Carinena}
  J.~F.~Carinena and M.~F.~Ranada,
\emph{``Comments on the presymplectic formalism and the theory of regular
  Lagrangians with constraints,''}
  J.\ Phys.\ A {\bf 28} (1995) L91,

\bibitem{Martinez}
S.~Martinez, J.~Cortes, M.~de Leon
\emph{``The geometrical theory of constraints applied to the dynamics of
vakonomic mechanical systems: The vakonomic bracket''}
J. Math. Phys. {\bf 41} (2000) 2090,

\bibitem{Lang}
S.~Lang \emph{``Differential and Riemannian Manifolds''} (1996) Springer.
\end{thebibliography}
\end{document}